\documentclass[a4paper,authoryear,5p,twocolumn,11pt]{elsarticle}
\pdfoutput=1

\usepackage{amsmath,amssymb}
\usepackage{booktabs}
\usepackage{url}
\usepackage{graphicx}
\usepackage{booktabs}
\usepackage{multirow}
\usepackage{capt-of}
\usepackage[utf8]{inputenc}
\usepackage{soul}
\usepackage{caption,subcaption}
\usepackage[normalem]{ulem}

\usepackage{color}
\definecolor{linkblue}{rgb}{0,0,.8}
\definecolor{linkgreen}{rgb}{0,0.45,0}
\definecolor{urlblue}{rgb}{0,0,0.9}
\definecolor{purple}{rgb}{0.7,0.0,0.4}
\usepackage[colorlinks,bookmarks]{hyperref}
\AtBeginDocument{
\hypersetup{linkcolor=linkblue,
            citecolor=purple,
            urlcolor=urlblue}}

\newcommand\be{\begin{equation}}
\newcommand\ee{\end{equation}}
\newcommand{\n}{\hat{\boldsymbol{n}}}
\newcommand{\uvec}{\boldsymbol{u}}
\newcommand{\vecl}[1]{\boldsymbol{v}_{#1}}

\usepackage[figuresright]{rotating}

\begin{document}

\begin{frontmatter}

\title{CMB statistical isotropy confirmation at all scales using multipole vectors}

\author[cca,ufes,uel]{Renan A. Oliveira}
\author[uel]{Thiago S. Pereira}
\author[ufrj,valongo]{Miguel Quartin}
\address[cca]{Center for Computational Astrophysics, Flatiron Institute, 10010-5902, New York, NY, USA}
\address[ufes]{PPGCosmo, CCE, Universidade Federal do Espírito Santo, 29075-910, Vitória, ES, Brazil}
\address[uel]{Departamento de Física, Universidade Estadual de Londrina, 86051-990, Londrina, PR, Brazil}
\address[ufrj]{Instituto de Física, Universidade Federal do Rio de Janeiro, 21941-972, Rio de Janeiro, RJ, Brazil}
\address[valongo]{Observatório do Valongo, Universidade Federal do Rio de Janeiro, 20080-090, Rio de Janeiro, RJ, Brazil}

\begin{abstract}
    We present an efficient numerical code and conduct, for the first time, a null and model-independent CMB test of statistical isotropy using Multipole Vectors (MVs) at all scales. Because MVs are insensitive to the angular power spectrum $C_\ell$, our results are independent from the assumed cosmological model. We avoid \emph{a posteriori} choices and use pre-defined ranges of scales $\ell\in[2,30]$, $\ell\in[2,600]$ and $\ell\in[2,1500]$ in our analyses. We find that all four masked Planck maps, from both 2015 and 2018 releases, are in agreement with statistical isotropy for $\ell\in[2,30]$, $\ell\in[2,600]$. For $\ell\in[2,1500]$ we detect anisotropies but this is indicative of simply the anisotropy in the noise: there is no anisotropy for $\ell < 1300$ and an increasing level of anisotropy at higher multipoles. Our findings of no large-scale anisotropies seem to be a consequence of avoiding \emph{a posteriori} statistics. We also find that the degree of anisotropy in the full sky (i.e.~unmasked) maps vary enormously (between less than 5 and over 1000 standard deviations) among the different mapmaking procedures and data releases.
\end{abstract}


\end{frontmatter}

\section{Introduction}

Cosmic Microwave Background (CMB) maps have been the best window to probe the hypotheses that the primordial perturbations were Gaussian and statistically homogeneous and isotropic. When these hypotheses are met, the multipolar coefficients of the CMB temperature map can be treated as random variables satisfying
\begin{equation}\label{cov-alm}
    \langle a_{\ell m}a^*_{\ell' m'}\rangle = C_\ell \delta_{\ell\ell'}\delta_{mm'}\,.
\end{equation}
where $C_\ell$ is the angular power spectrum and $\delta_{ij}$ is the Kronecker delta. CMB experiments have spent the last decades in pursuit of a precise measurement of the $C_\ell$s. The WMAP mission successfully measured this quantity to the cosmic variance limit in the range $2\leq\ell \leq 600$, showing a remarkable accuracy between theory and observations~\citep{Hinshaw:2012aka,Bennett:2012zja}. The Planck team then extended this task to the multipole range $2\leq\ell\leq 1800$, confirming the predictions of the standard model with unprecedented precision~\citep{Akrami:2018vks,Aghanim:2018eyx}.

In the standard framework, the $C_\ell$s are, at each $\ell$, the variance of the distribution from which the primordial perturbations were drawn, and at first order in perturbation theory constitute both the \emph{only} non-trivial statistical moment of a CMB map and the \emph{only} quantity predicted by theory. Since each multipole has only $2\ell+1$ independent components, this imposes a fundamental lower-bound to the uncertainty in measuring the $C_\ell$s, known as \emph{cosmic variance}. This means that in a typical Planck map, the over $3$ million modes ($a_{\ell m}$s) measured in the cosmic variance limit (apart from the masked regions) are reduced to only 1800 numbers ($C_\ell$s): a data reduction of a factor of almost 2000.

Most fundamental extensions of the standard cosmological model will modify equation~{\eqref{cov-alm}}, either by including extra off-diagonal terms and/or by introducing higher order correlations between the $a_{\ell m}$s.\footnote{Some models will also change the $C_\ell$s while keeping the matrix~\eqref{cov-alm} diagonal. But as we will see, MVs are insensitive to such models.} The challenge is that most statistical estimators built to extract such extensions will compress the $a_{\ell m}$s in a model~\citep{Aghanim:2013suk,Amendola:2010ty,Prunet:2004zy} or geometrical~\citep{Pullen:2007tu,Froes:2015hva,Hajian:2004zn} dependent way, and thus meaningful information could still remain undetected.

At the same time, there have been claims of possible anomalies in the CMB data~\citep[see][for a review]{Ade:2015hxq,Schwarz:2015cma,Muir:2018hjv}. These claims are hard to verify because they are mostly related to large-scales which have already been measured at the cosmic variance limit since WMAP. Thus one cannot settle the issue with just more observations of the same quantities.\footnote{Although one could expect to see similar effects in the polarization data.} For instance, for the quadrupole--octupole alignment~\citep{Tegmark:2003ve,Bielewicz:2005zu,Copi:2005ff,Abramo:2006gw} most recent papers focused
either on the study of possible systematics~\citep{Francis:2009pt,Rassat:2014yna,Notari:2015kla}, or on the signatures of alignments using galaxy catalogs~\citep{Tiwari:2018hrs}. Another issue with the study of anomalies is how to deal with the so-called \emph{look elsewhere} effect: \emph{a posteriori} selection of a small subset of the current 3 million modes of the CMB can artificially lead to low-probability statistics~\citep{Bennett:2010jb}.

In this work we circumvent these issues is by making use of Multipole Vectors (MVs)~\citep{Copi:2003kt,Katz:2004nj}. MVs are a well-known alternative decomposition for multipole fields~\citep{maxwell1873treatise,Castillo:2004}, and are particularly suited to the analysis of CMB maps, since they allow for model-independent tests of isotropy. We also introduce a novel set of vectors, dubbed as Fréchet Vectors (FVs), which are defined as the mean position of a set of MVs on the sphere.  Not only the tools that we use are motivated~\emph{a priori}, but we also choose to work on three well motivated range of scales.

The advantage of the FVs is that, besides sharing most of the nice properties of the MVs, they capture the correlation of the latter in a model-independent way.

\section{Multipole Vectors and Fréchet Vectors}

MVs are an alternative representation to the multipole moments of functions on the sphere. For any function $X(\n)$, its multipole moments  ${X_\ell(\n)=\sum_m X_{\ell m}Y_{\ell m}(\n)}$ can be specified in terms of a real constant $\lambda_\ell$ and $\ell$ unit and headless (multipole) vectors $\vecl{\ell}$ as
\begin{equation}
    X_\ell (\n) = \left.\lambda_\ell \nabla_{\vecl{1}}\cdots\nabla_{\vecl{\ell}}\frac{1}{r}\right|_{r=1} \,,
\end{equation}
where $r=\sqrt{x^2+y^2+z^2}$ and $\nabla_{\vecl{\ell}} = \vecl{\ell}\cdot\nabla$. In the case of CMB, $X$ could be either the temperature ($T$) or polarization ($E$ or $B$-modes) fluctuations. Being vectors, MVs do not depend on external frames of reference, but instead rotate rigidly with the data. Moreover, all the information on $C_\ell$ is contained in $\lambda_\ell$ (see below), so the vectors do not depend on cosmology in the standard, Gaussian FLRW case. It is thus natural to think about the $a_{\ell m}$s as represented by the $2\ell+1$ numbers of the set $\{C_\ell,\vecl{1},\cdots, \vecl{\ell}\}$~\citep{Copi:2003kt}.

Previous CMB analysis using MVs have focused mostly on the low range of scales, $2\leq \ell \lesssim 50$, and were mostly interested in their power to detect large angle statistical anomalies~\citep{Copi:2003kt,Schwarz:2004gk,Land:2005ad,Bielewicz:2005zu,Abramo:2006gw,Bielewicz:2008ga,Pinkwart:2018nkc}. Here we will use these vectors to conduct, for the first time, a null test of statistical isotropy in the range $2\leq\ell\leq 1500$. Multipoles ${\ell\gtrsim 1500}$ are affected by the anisotropic instrumental noise~\citep{Adam:2015tpy}, so their inclusion is postponed to a future analysis.

Algorithms for extracting MVs from the $a_{\ell m}$s were given in~\citet{Copi:2003kt,Weeks:2004cz}~\citep[see][for a recent review on the existing algorithms]{Pinkwart:2018nkc}. A much more elegant and faster algorithm was given in~\citep{Helling:2006xh,dennis2004canonical,Dennis:2004sz}, and is based in the fact that MVs can be identified with the roots of a
random polynomial $Q_{\ell}$ having the $a_{\ell m}$s as coefficients:
\begin{equation}\label{Q-poly}
Q_\ell(z) = \sum_{m=-\ell}^{\ell}\sqrt{\binom{2\ell}{\ell+m}}\, a_{\ell m}\, z^{\ell+m}\,.
\end{equation}
For each $\ell$, this polynomial has $2\ell$ complex roots $z_i$, $i=\{1,\cdots,2\ell\}$. However, only half of these are independent, since the other half can be obtained by the relation $z \rightarrow -1/z^*$.\footnote{This reflects the parity invariance of the MVs, which is ultimately linked to the reality of the CMB field.} Given a root $z_i$, one can obtain the pair $(\theta_i,\phi_i)$ of coordinates of the vector by means of a stereographic projection ${z_i = \cot(\theta_i/2)e^{{\rm i}\phi_i}}$.

We have built a new Python code dubbed \texttt{polyMV} which uses \texttt{MPSolve}~\citep{bini2014solving} to find the roots of Eq.~\eqref{Q-poly} and convert a set of $z_i$s into a set of $(\theta_i,\phi_i)$ coordinates. Computational time tests in obtaining all MVs at a given $\ell$ show that our code has computational complexity ${\cal O}(\ell^2)$, compared to ${\cal O}(\ell^{3.5})$ of the ones in~\citep{Copi:2003kt,Weeks:2004cz}. See~\ref{app:timings} for more details. This means that in a simple 2015 desktop it takes less than 1 sec to extract all MVs at $\ell=1000$, compared to around $75$ min and $22$~h with the routines of~\cite{Copi:2003kt} and~\cite{Weeks:2004cz}, respectively.  Our code comparison also served as a cross-check: the absolute difference  between our MV values and those of~\cite{Copi:2003kt} was	only $\sim 10^{-10}$.

The independence of the MVs on the $C_\ell$ can be directly seen in~\eqref{Q-poly}. By rescaling the $a_{\ell m}$s as
${a_{\ell m} = \sqrt{C_\ell} b_{\ell m}}$, with ${\langle b_{\ell m}b^*_{\ell' m'}\rangle =\delta_{\ell\ell'} \delta_{mm'}}$, one obtains an equivalent class of polynomials $R_\ell(z)$, all having the same roots as $Q_\ell(z)$. For this reason the addition of Gaussian and {\it isotropic} noise to a CMB map will not change the statistics of the MVs, since the sum of Gaussian and isotropic variables is still Gaussian and isotropic and the net effect will be just a change on the effective $C_\ell$s.

If the $a_{\ell m}$s are drawn from a Gaussian, isotropic and unmasked random sky, the one-point function of the roots
$z_i$ (i.e., the expected values $\langle z_i \rangle$) follow a uniform distribution on the Riemann
sphere~\citep{bogomolny1992prl}, so that the normalized\footnote{This normalization uses the fact
	that MVs are headless, so that we only consider vectors in the upper hemisphere.} one-point function of the stereographic angles are
${P_1^\ell(\theta,\phi)=(\sin\theta d\theta)(d\phi/2\pi)}$. In terms of the variables
\begin{equation}\label{eq:eta-phi}
\eta\,\equiv\, 1-\cos\theta \qquad {\rm and} \qquad \varphi \,\equiv\, \phi/2\pi
\end{equation}
this reduces to
\begin{equation}\label{opf-mvs}
P_1^\ell(\eta,\varphi)=d\eta d\varphi\times
\begin{cases}
	1 & (\eta,\varphi)\in[0,1]\,,\\
	0 & \text{otherwise}\,.
\end{cases}
\end{equation}

Moreover, MVs at different multipoles are uncorrelated whenever the $a_{\ell m}$s are. However, vectors coming from the same multipole \emph{are} correlated, and will in general have all $N$-point correlation functions, with ${1\leq N \leq\ell}$~\citep{dennis2004canonical,Dennis:2004sz}. Thus, in order to scrutinize CMB data against the null statistical hypothesis, it is interesting to test not only Eq.~\eqref{opf-mvs}, but also the inner correlations of MVs at a fixed multipole. Testing the uniformity of MVs is straightforward as we shall see in the next section. Testing their inner correlations, on the other hand, is a much harder task, given that the number of possible correlations grows as $\ell^2$. This motivates us to look for ways of mapping the MV vectors into a smaller set of vectors. While there have been similar approaches in the literature~\citep[see e.g.][]{Abramo:2006gw,Copi:2003kt}, they were designed to test for specific CMB anomalies, which is not our primary goal. That is to say, we would like to compress the information in the MVs without introducing biases towards specific signatures.

\begin{figure*}[t!]
    \centering 
\begin{subfigure}{0.25\textwidth}
  \includegraphics[scale=0.17]{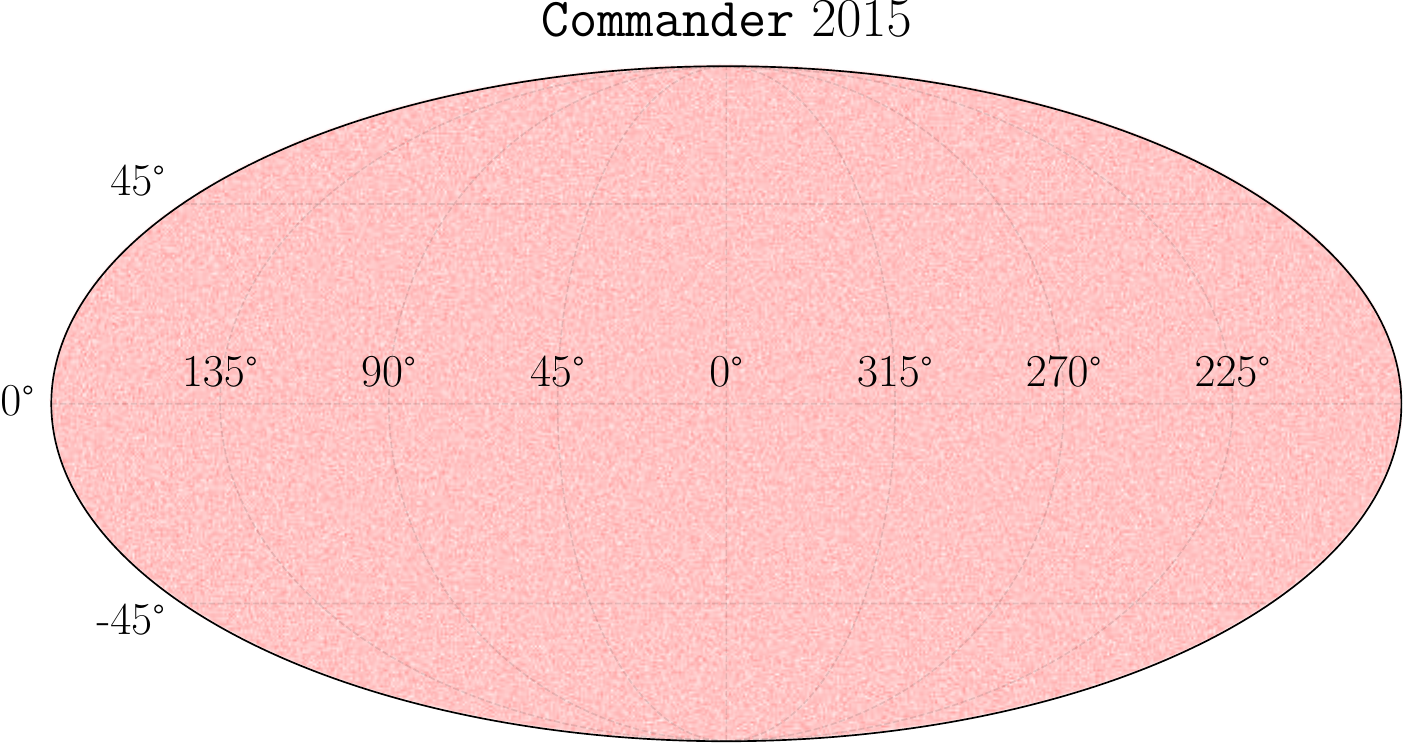}
\end{subfigure}\hfil 
\begin{subfigure}{0.25\textwidth}
  \includegraphics[scale=0.17]{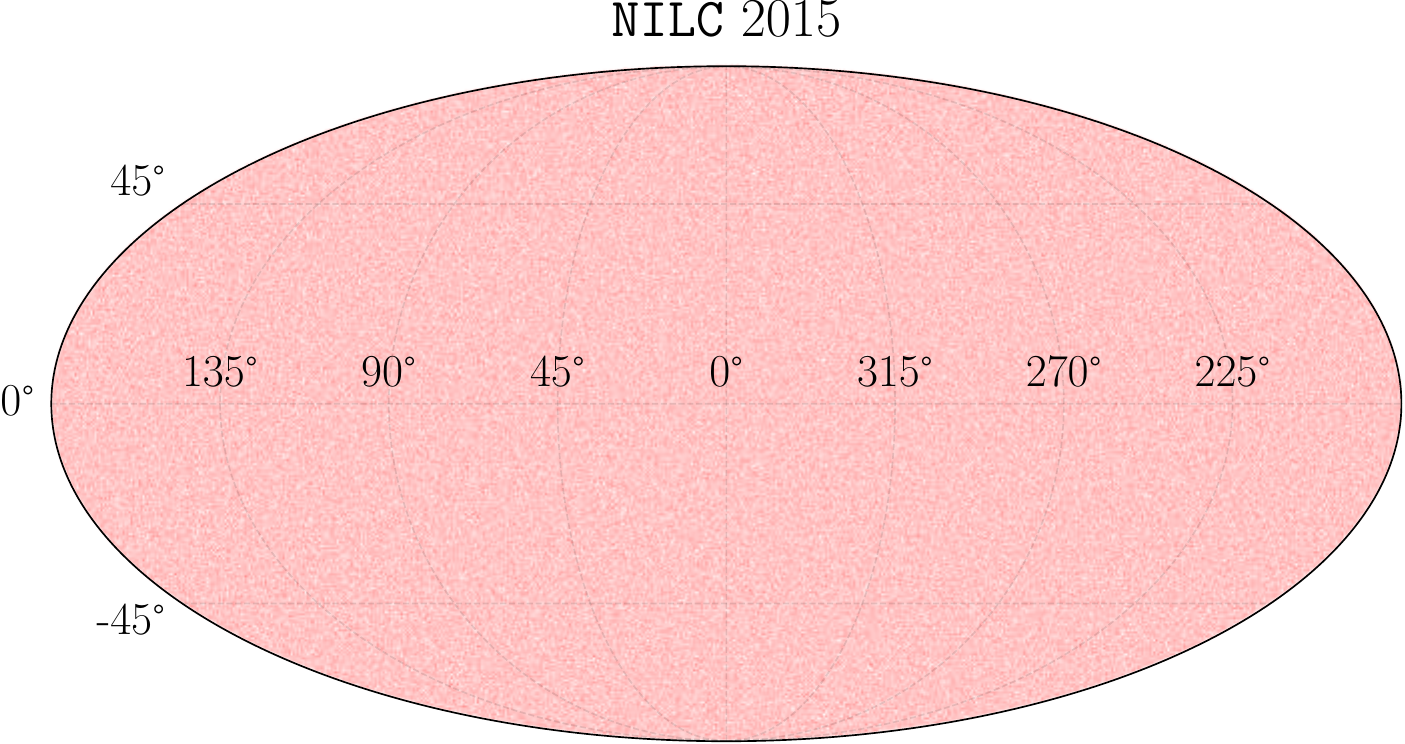}
\end{subfigure}\hfil 
\begin{subfigure}{0.25\textwidth}
  \includegraphics[scale=0.17]{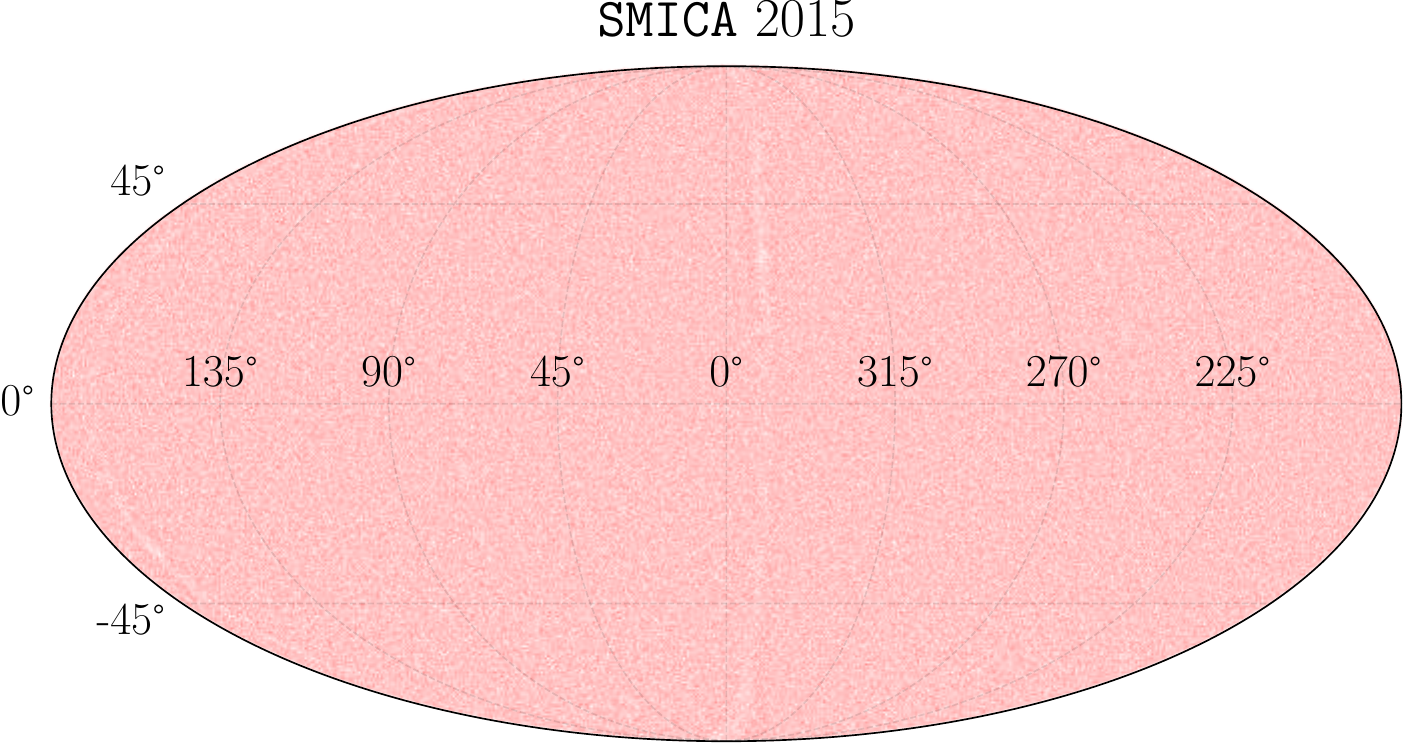}
\end{subfigure}\hfil
\begin{subfigure}{0.25\textwidth}
  \includegraphics[scale=0.17]{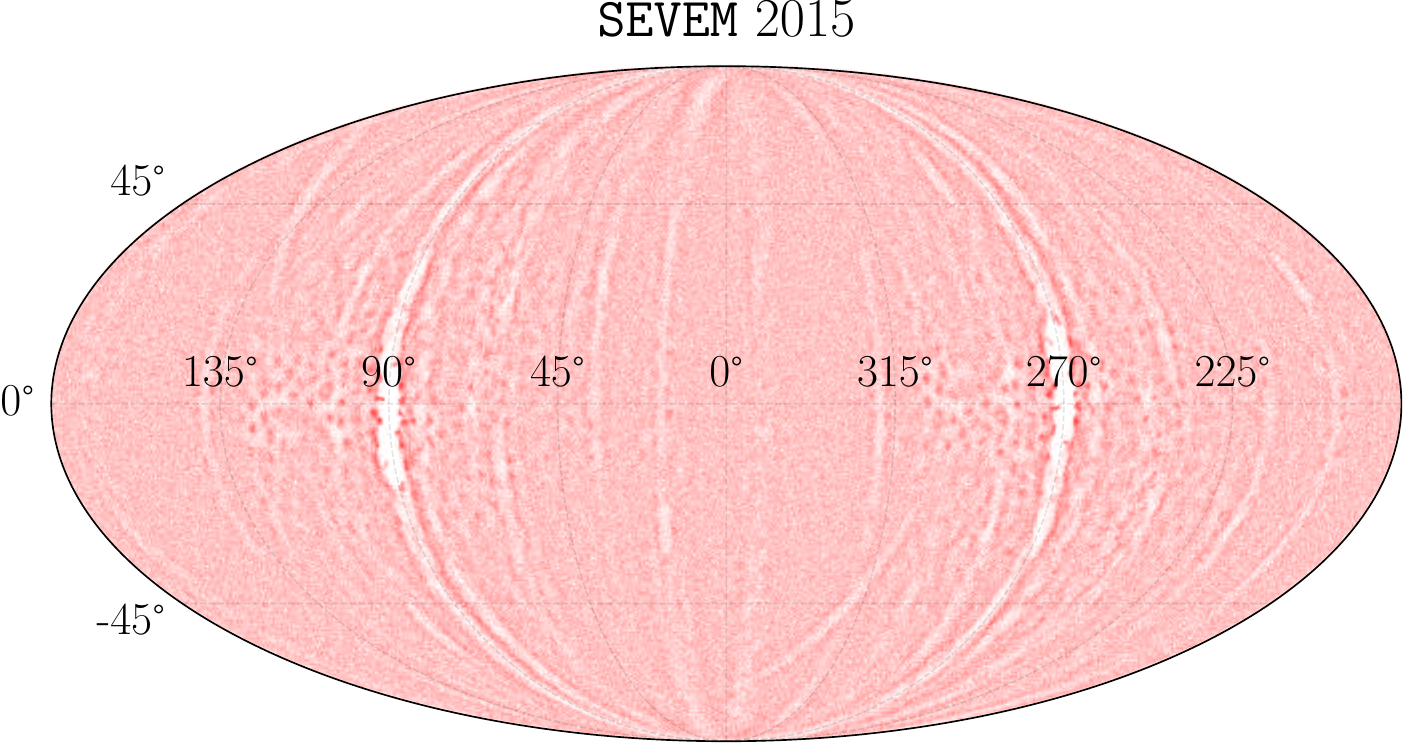}
\end{subfigure}

\medskip
\begin{subfigure}{0.25\textwidth}
  \includegraphics[scale=0.17]{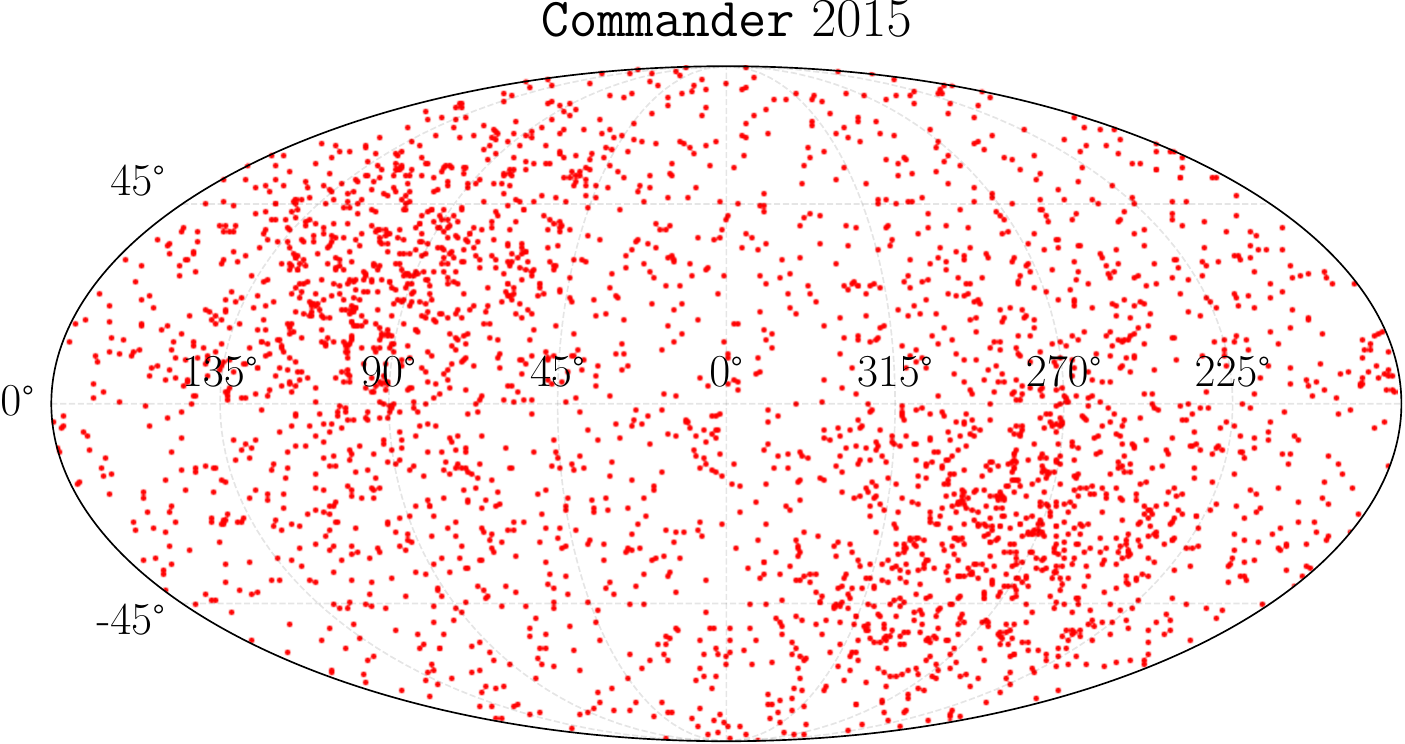}
\end{subfigure}\hfil 
\begin{subfigure}{0.25\textwidth}
  \includegraphics[scale=0.17]{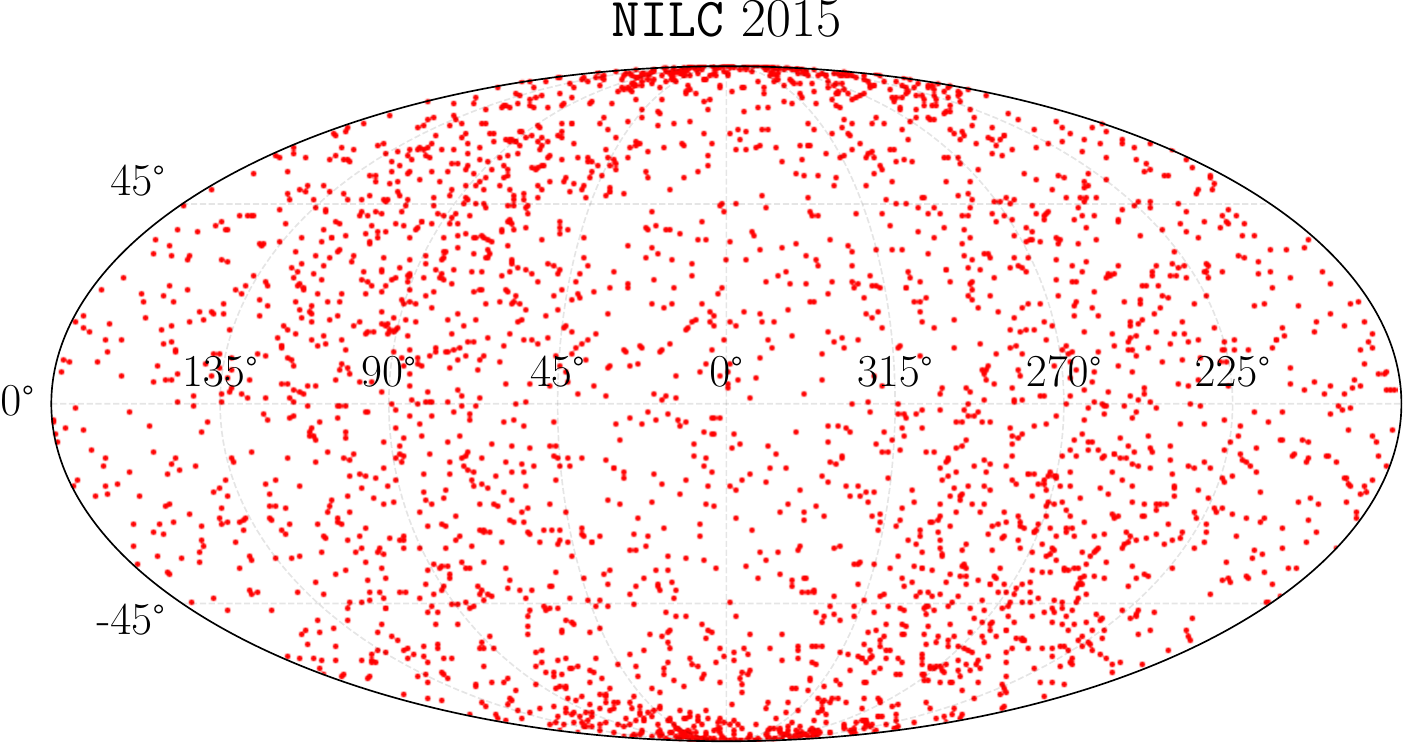}
\end{subfigure}\hfil 
\begin{subfigure}{0.25\textwidth}
  \includegraphics[scale=0.17]{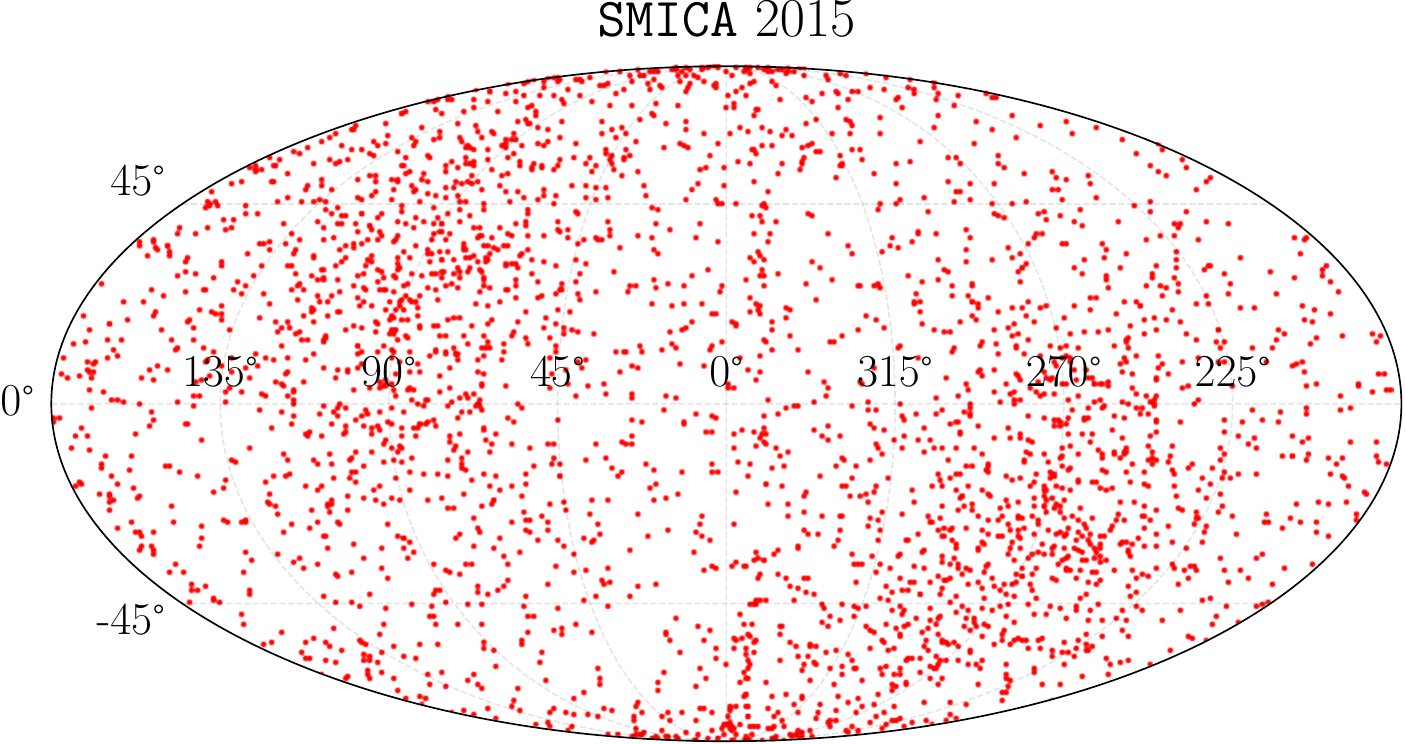}
\end{subfigure}\hfil
\begin{subfigure}{0.25\textwidth}
  \includegraphics[scale=0.17]{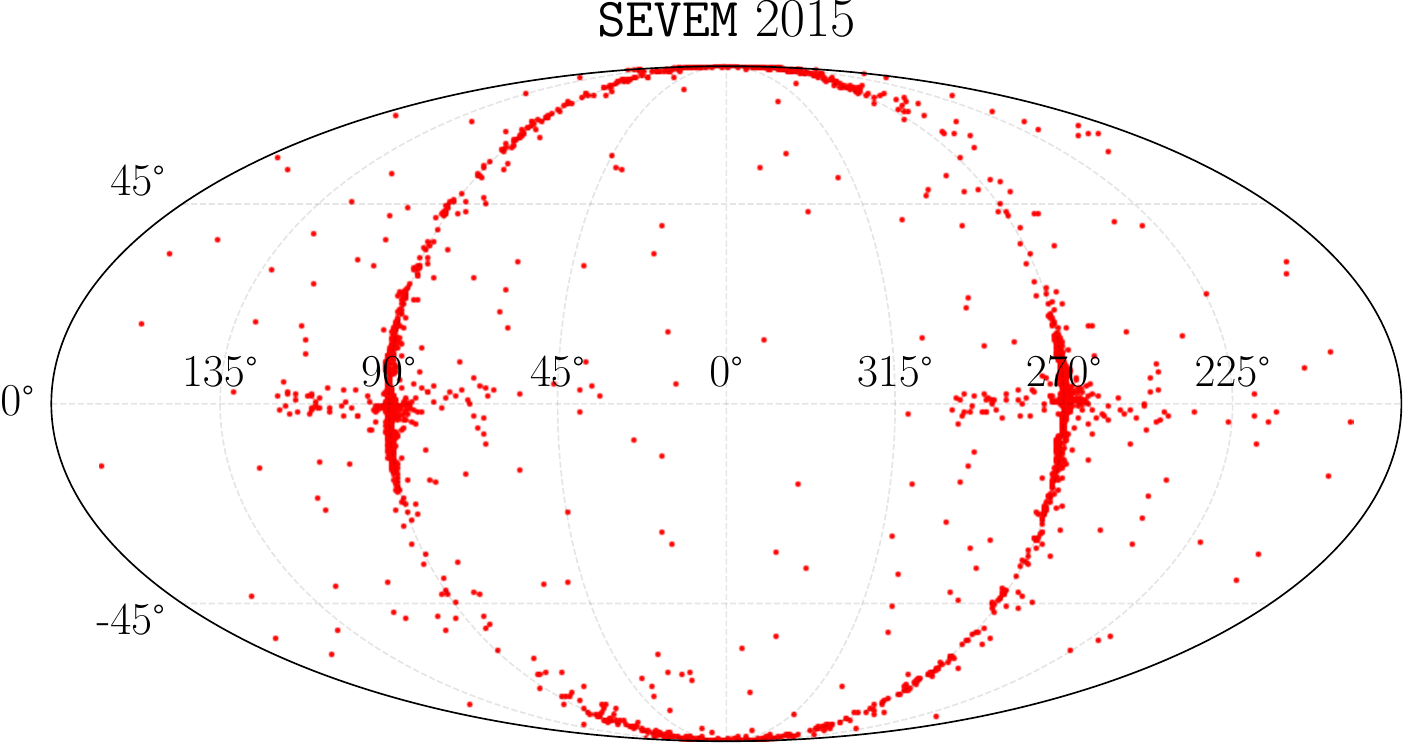}
\end{subfigure}

\medskip
\begin{subfigure}{0.25\textwidth}
  \includegraphics[scale=0.17]{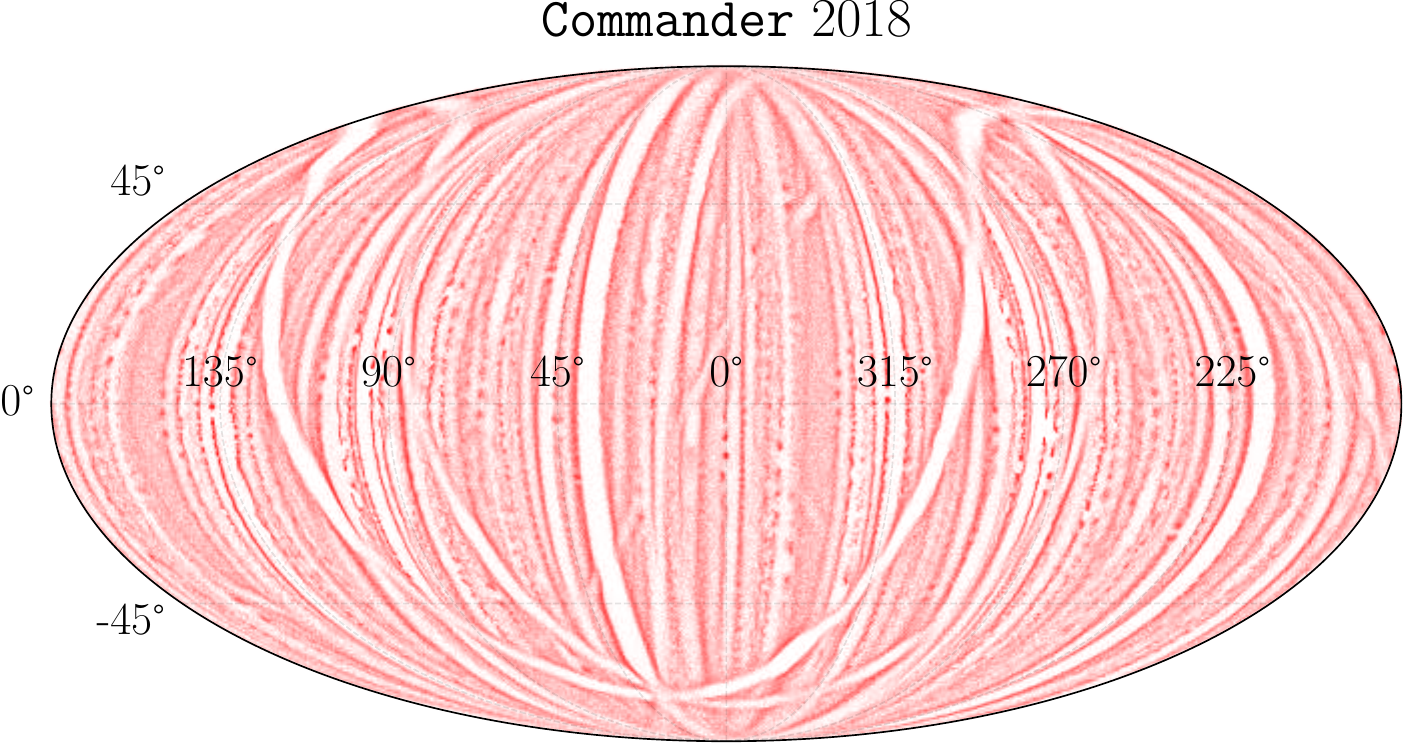}
\end{subfigure}\hfil 
\begin{subfigure}{0.25\textwidth}
  \includegraphics[scale=0.17]{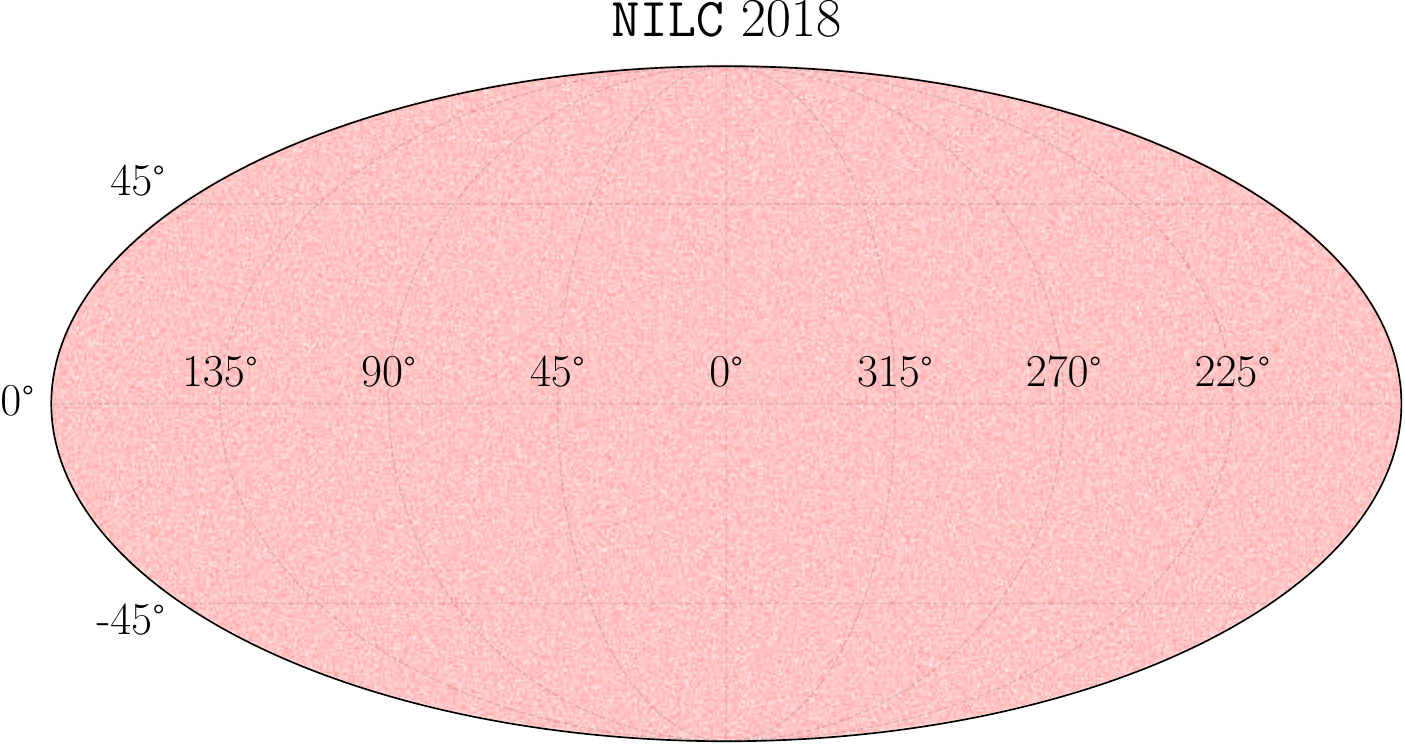}
\end{subfigure}\hfil 
\begin{subfigure}{0.25\textwidth}
  \includegraphics[scale=0.17]{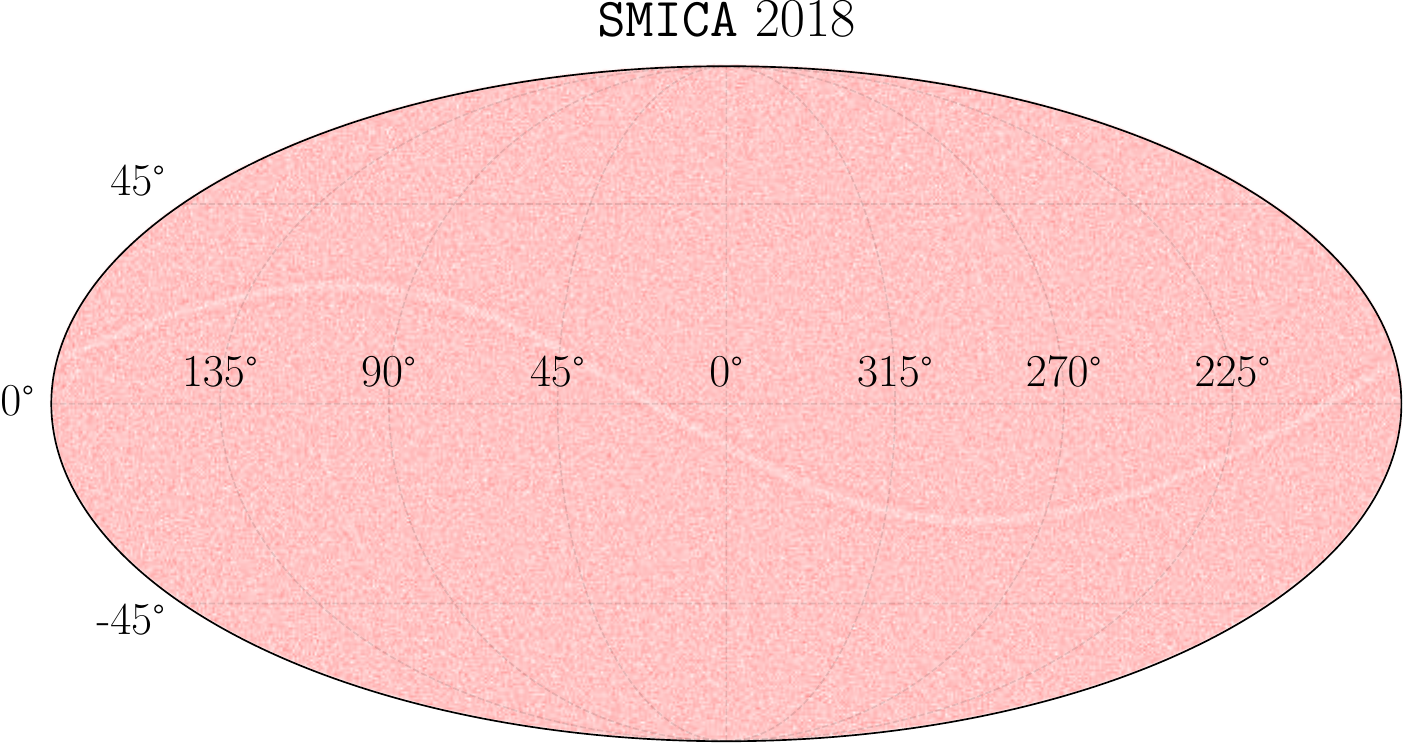}
\end{subfigure}\hfil
\begin{subfigure}{0.25\textwidth}
  \includegraphics[scale=0.17]{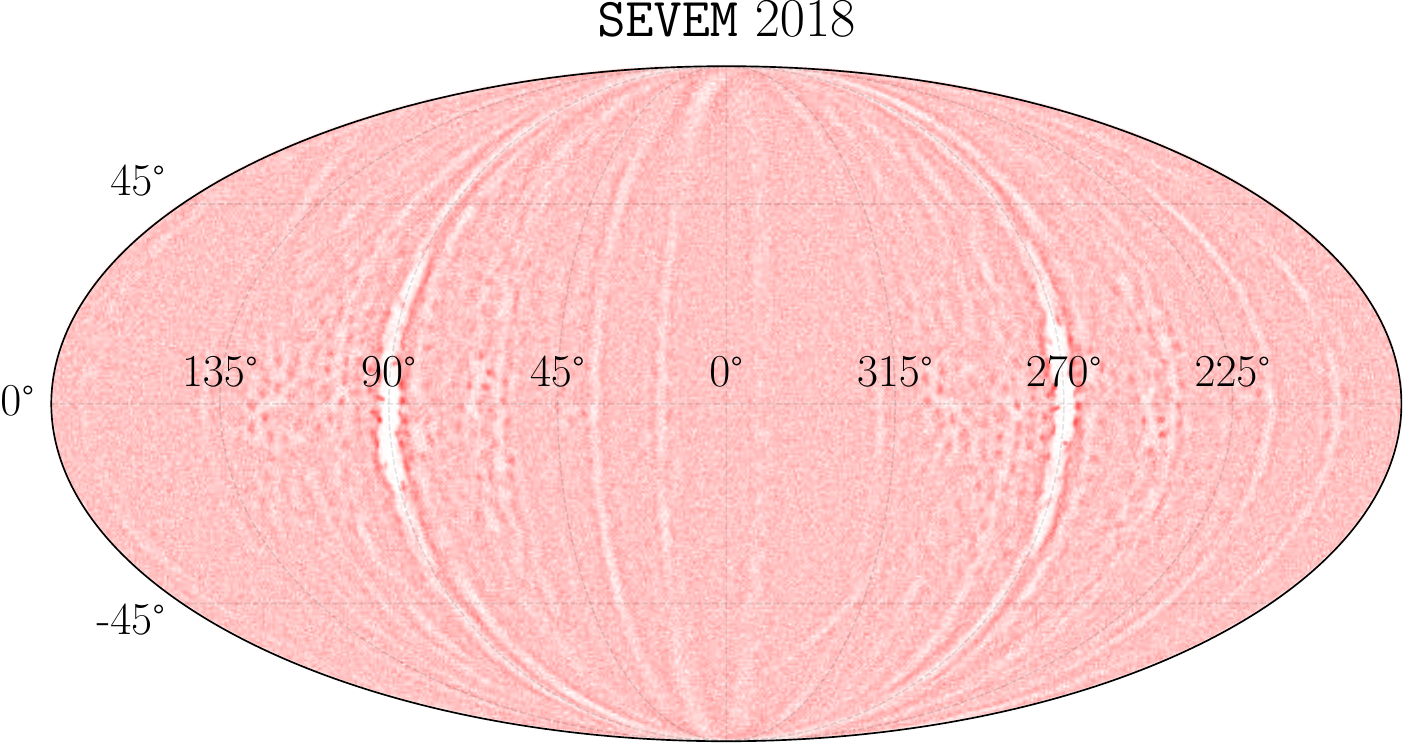}
\end{subfigure}

\medskip
\begin{subfigure}{0.25\textwidth}
  \includegraphics[scale=0.17]{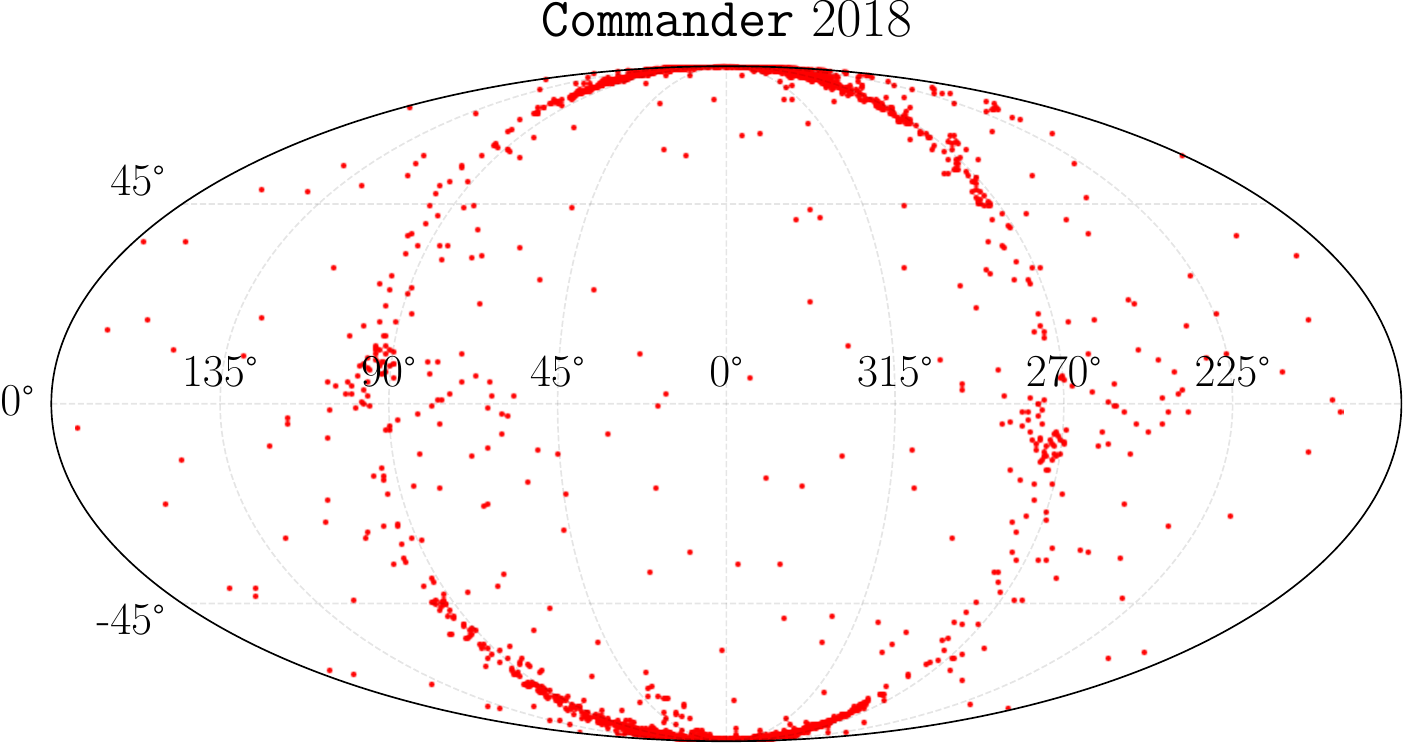}
\end{subfigure}\hfil 
\begin{subfigure}{0.25\textwidth}
  \includegraphics[scale=0.17]{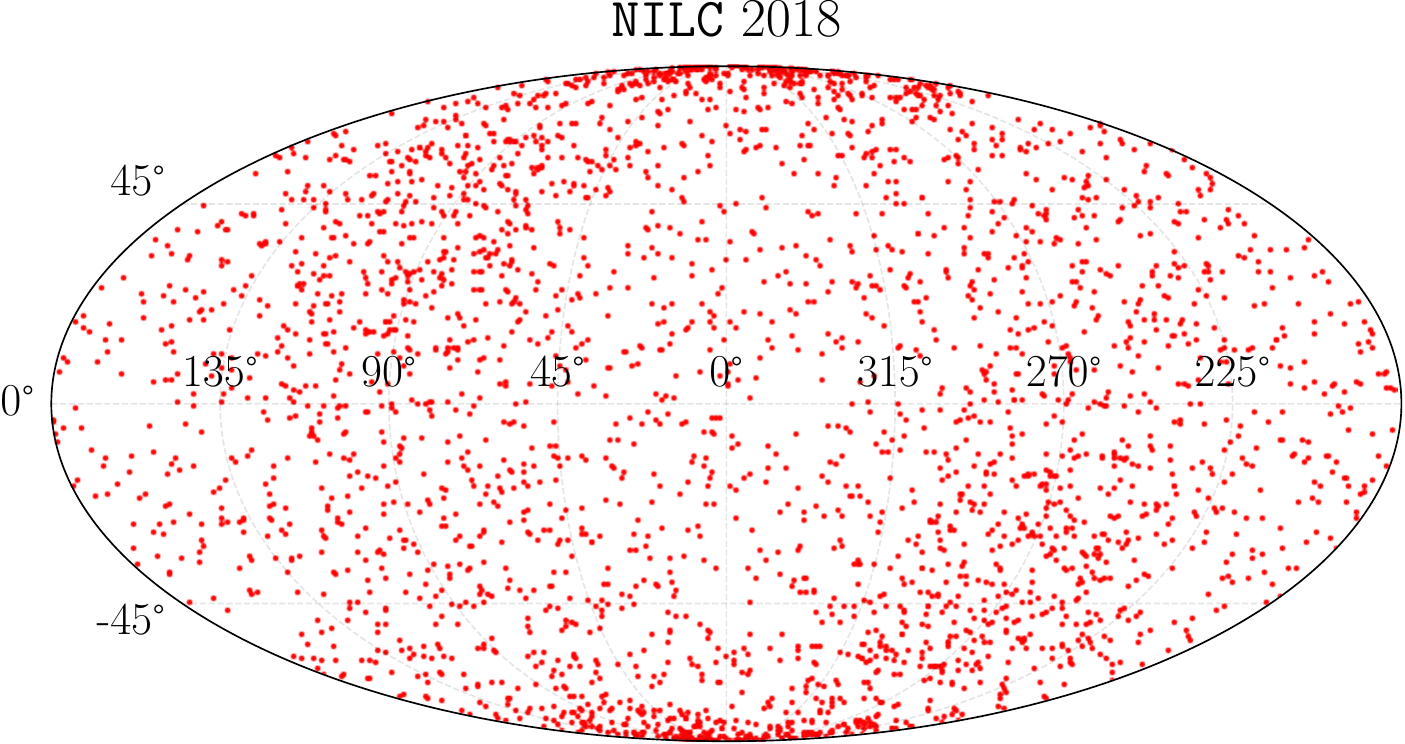}
\end{subfigure}\hfil 
\begin{subfigure}{0.25\textwidth}
  \includegraphics[scale=0.17]{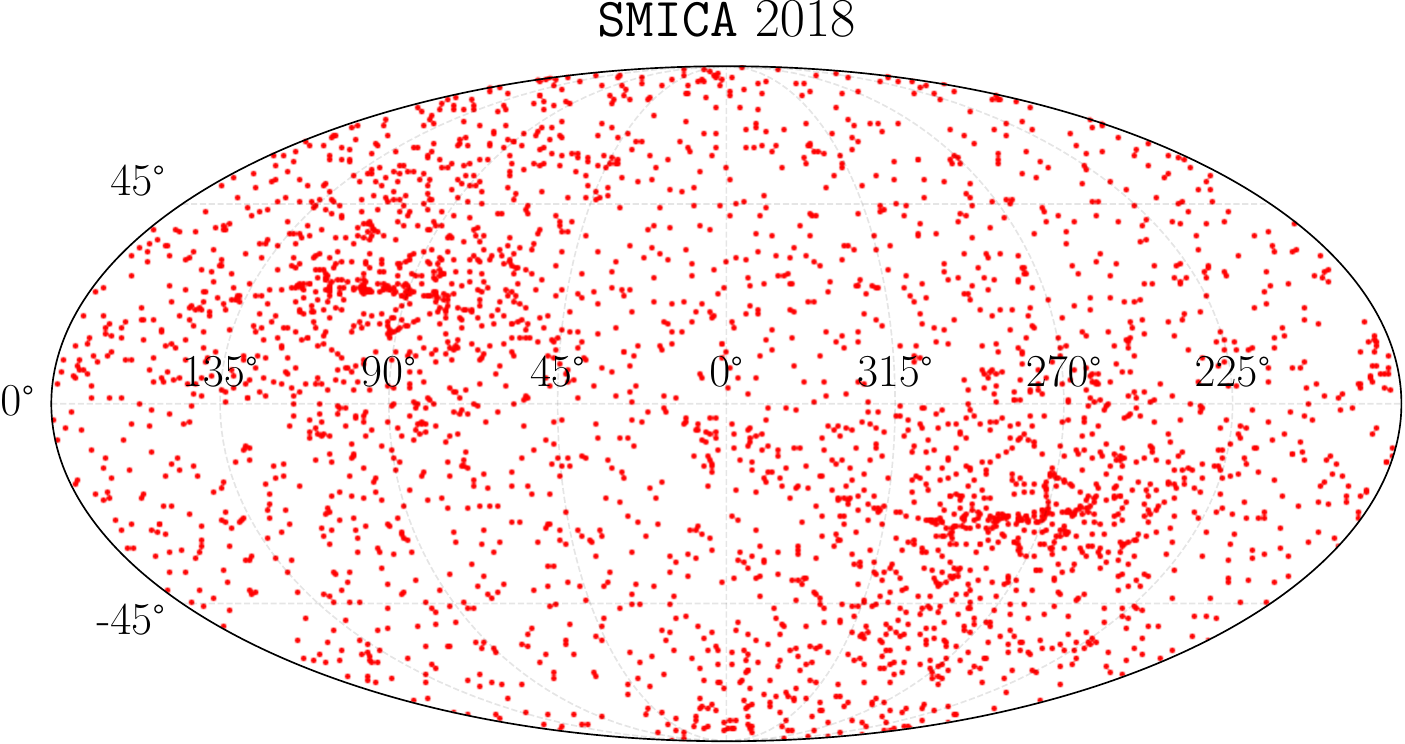}
\end{subfigure}\hfil
\begin{subfigure}{0.25\textwidth}
  \includegraphics[scale=0.17]{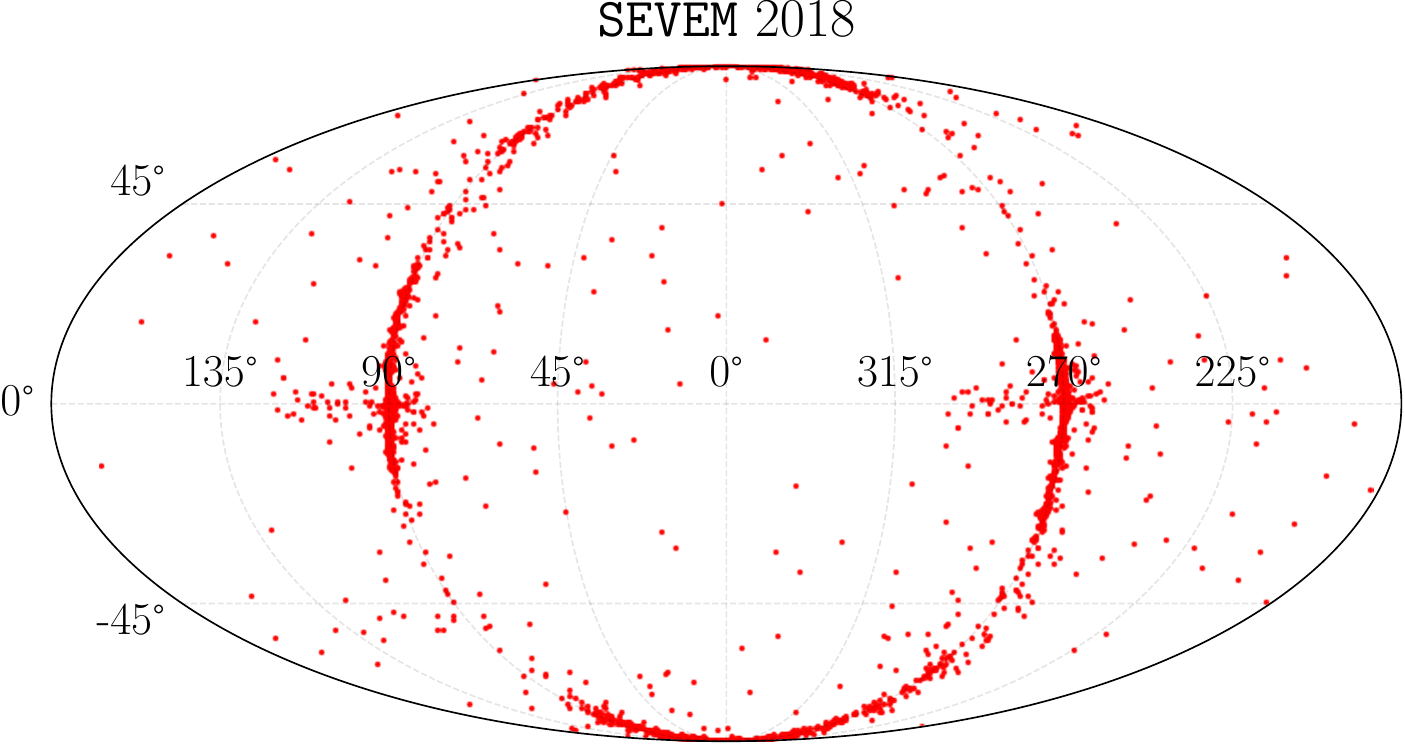}
\end{subfigure}
\caption{
Plot of MVs and FVs in galactic coordinates in the range $\ell\in[2,1500]$. Each pair of (antipodal) red dots represents one MV or FV.
\emph{First row:} MVs  for the unmasked Planck 2015 maps.
\emph{Second row:} corresponding FVs.
\emph{Third and fourth rows:} same for the unmasked Planck 2018 maps.
Note that 3 out of the 8 MV unmasked maps appear clearly anisotropic: \texttt{SEVEM} 2015 and 2018, and \texttt{Commander} 2018. Note also that FVs appear anisotropic even for MV sets which, by eye, look isotropic, such as \texttt{Commander} 2015 and \texttt{NILC} 2015 and 2018.
}
\label{mollweide-mvs-fvs-fullsky}
\end{figure*}

One of the simplest ways is to separate the MVs according to multipole, and analyse separately the $\ell$ MVs in each multipole $\ell$. If we think of MVs as points on the unit sphere, one possibility is to look for the point which minimizes the sum of squared distances between itself and all other points. That is, we can look for the vector $\uvec_{{\rm min},\ell}$ such that
\begin{equation}
    \uvec_{{\rm min},\ell} = {\rm arg\, min\,} \Psi_{\ell}\,,
\end{equation}
where
\begin{equation}
    \Psi_{\ell}(\uvec) = \sum_{i=1}^{2\ell} \gamma^2(\uvec,\vecl{i,\ell})\,,\quad{\textrm{with}}\quad \cos\gamma = \uvec\cdot\vecl{i,\ell}\,.
\end{equation}
The function $\Psi_{\ell}$ is a generalization of the variance for data points living on the sphere, and is known as the Fréchet variance. For this reason we shall dub $\uvec_{{\rm min},\ell}$ as the \emph{Fréchet Vectors} (FV).\footnote{In the mathematics literature it is known as Fréchet mean, or sometimes as Riemannian center of mass~\citep{nielsen2013matrix}.} Fréchet vectors have many interesting properties that we shall explore in a future publication. For our present purposes, it suffices to say that, just like the MVs, they are headless and their one-point function is uniform under the null hypothesis; thus, testing for their uniformity is also straightforward. As we shall see statistics on FVs can be more sensitive than similar statistics on the MVs.

The presence of masks will change the statistics of both Multipole and Fréchet vectors, and we resort to numerical simulations to estimate their distributions. In this work we will focus on the Planck temperature maps in the range $2\leq\ell\leq 1500$, where each mode is measured with a high signal-to-noise (S/N) ratio. We will not consider polarization in this first paper as both $E$ and $B$-modes have (individually) low S/N and Planck has a highly anisotropic noise profile. This would lead to strong noise-induced anisotropies and we would need to take into account the Planck noise simulations. The Planck team provided, in each release, 4 different temperature maps: \texttt{Commander}, \texttt{NILC}, \texttt{SEVEM} and \texttt{SMICA}. Each is built using a different pipeline but all use as input the different intensity frequency maps and aim at removing all foregrounds as much as possible. Figure~\ref{mollweide-mvs-fvs-fullsky} shows all MVs and corresponding FVs for these four full sky (unmasked) 2015 and 2018 Planck maps.\footnote{The \texttt{polyMV} code is available at~\url{https://oliveirara.github.io/polyMV/}, and the tables of the MVs and FVs for all the Planck maps considered in this work are available at~\url{https://doi.org/10.5281/zenodo.3866410}.}

\begin{figure*}[h!]
    \centering 
    \includegraphics[scale=0.22]{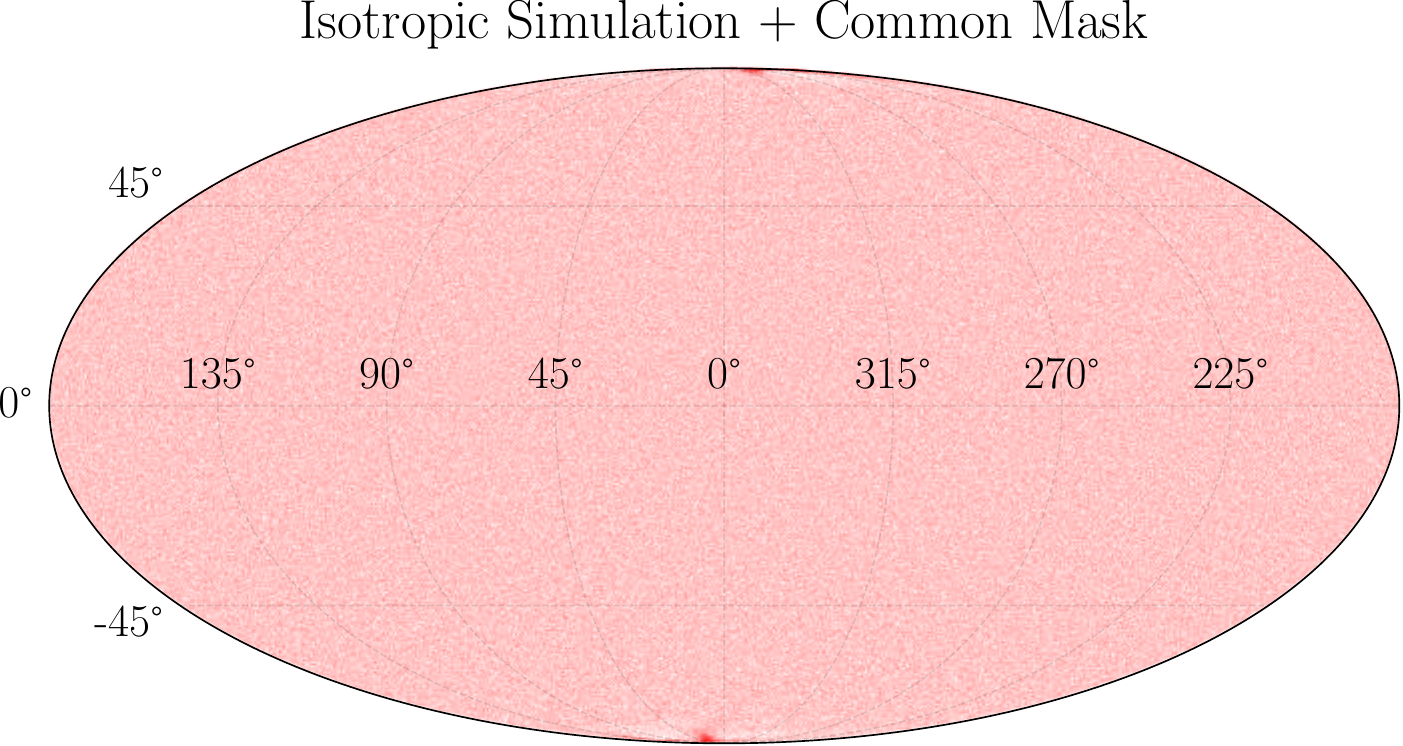}
    \includegraphics[scale=0.22]{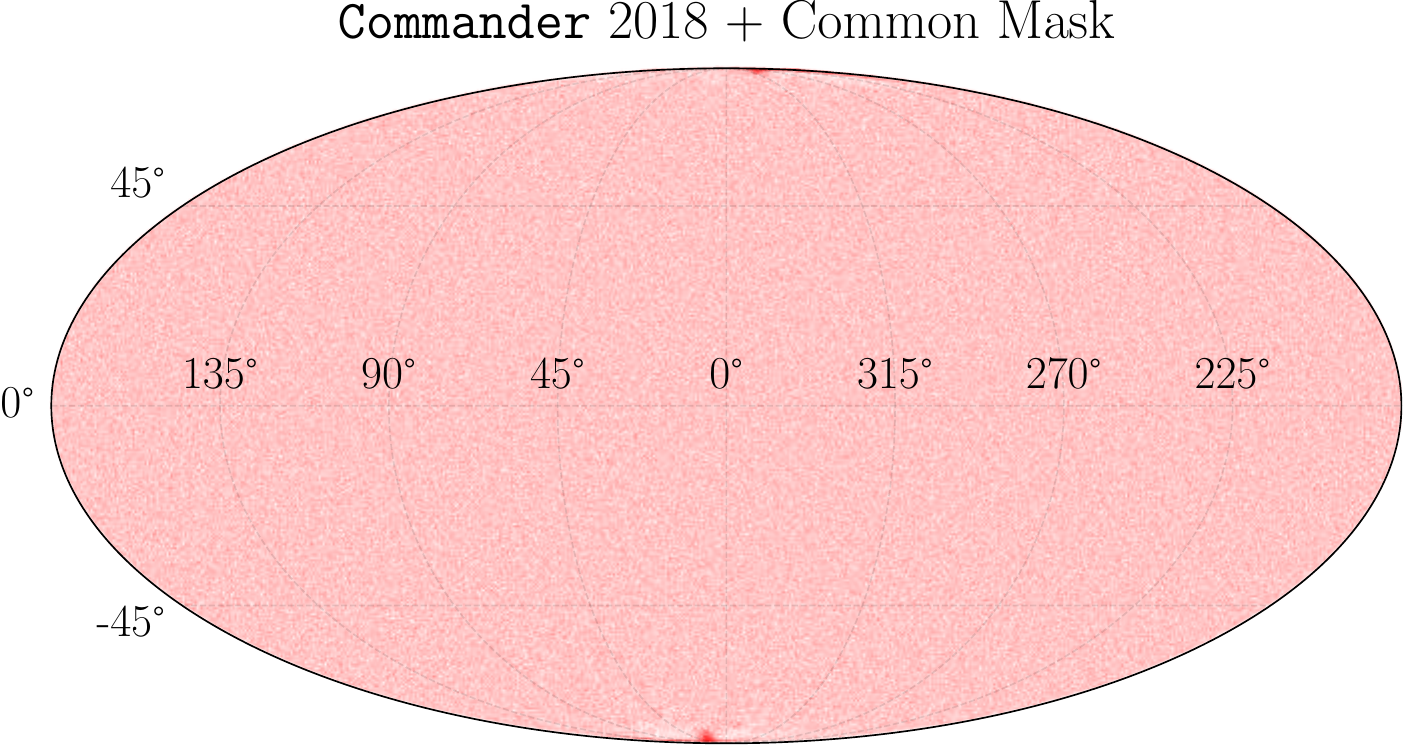}
    \includegraphics[scale=0.22]{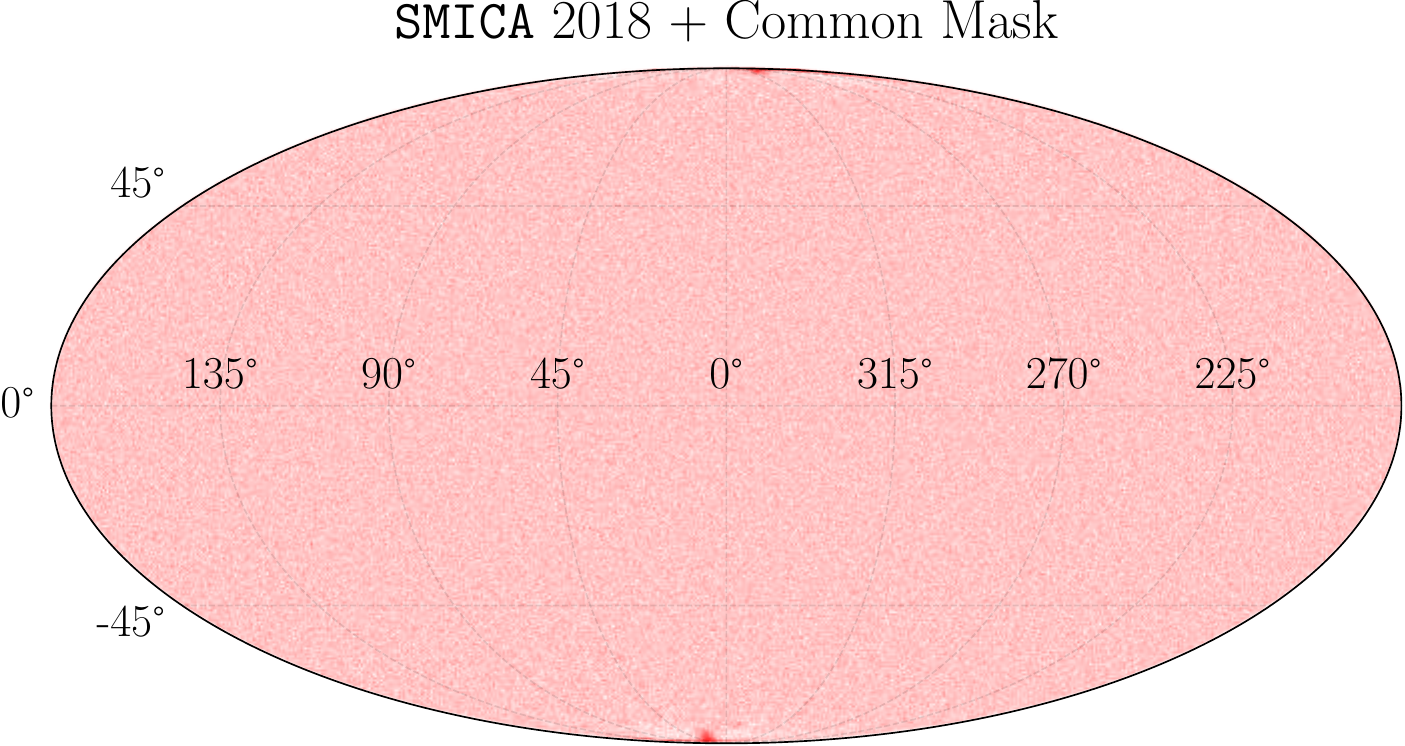}
    \includegraphics[scale=0.22]{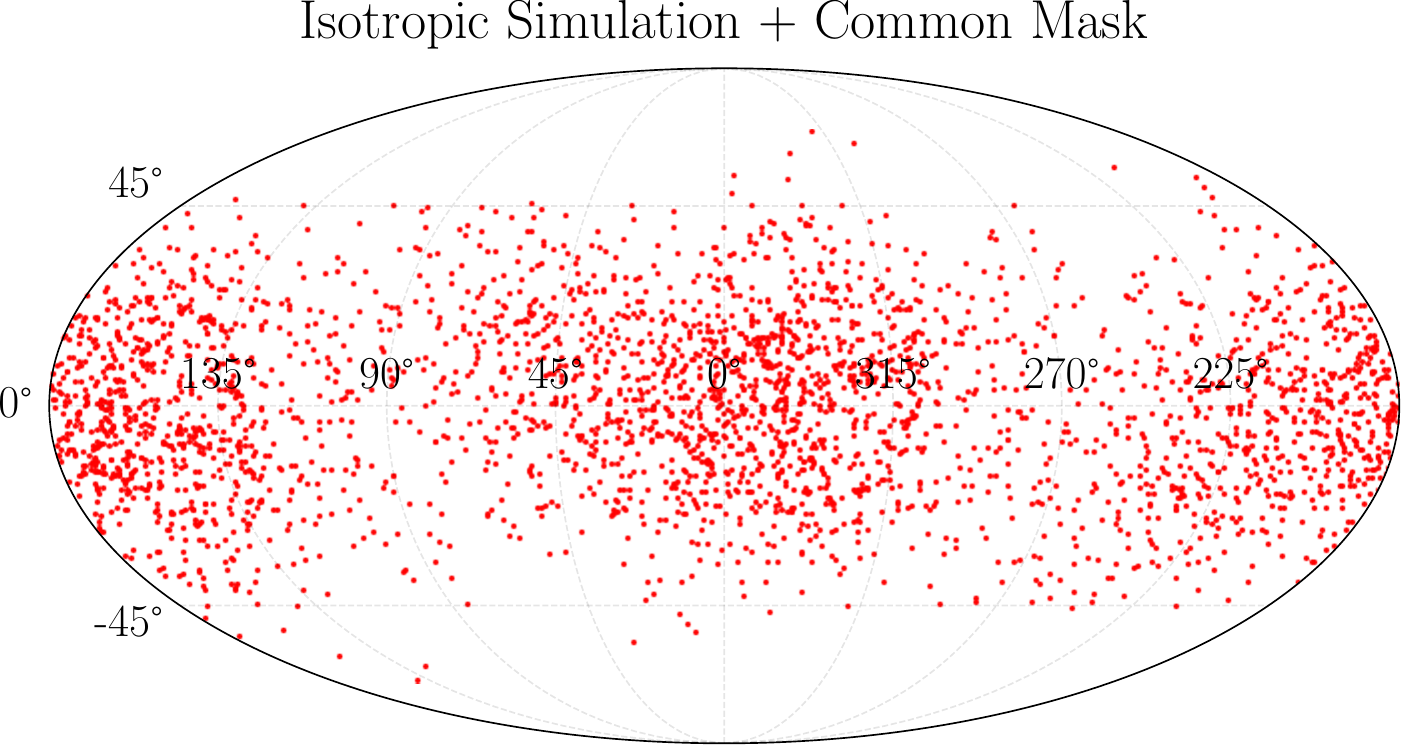}
    \includegraphics[scale=0.22]{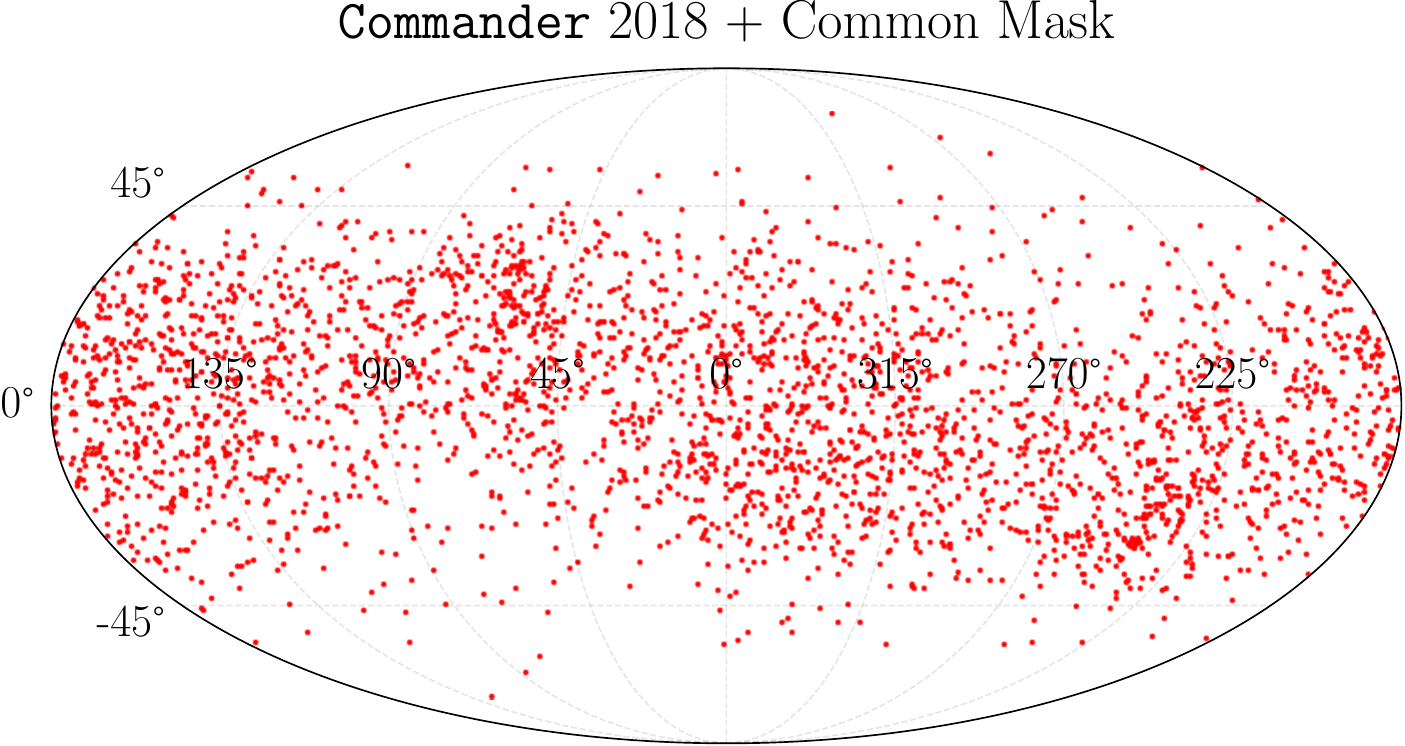}
    \includegraphics[scale=0.22]{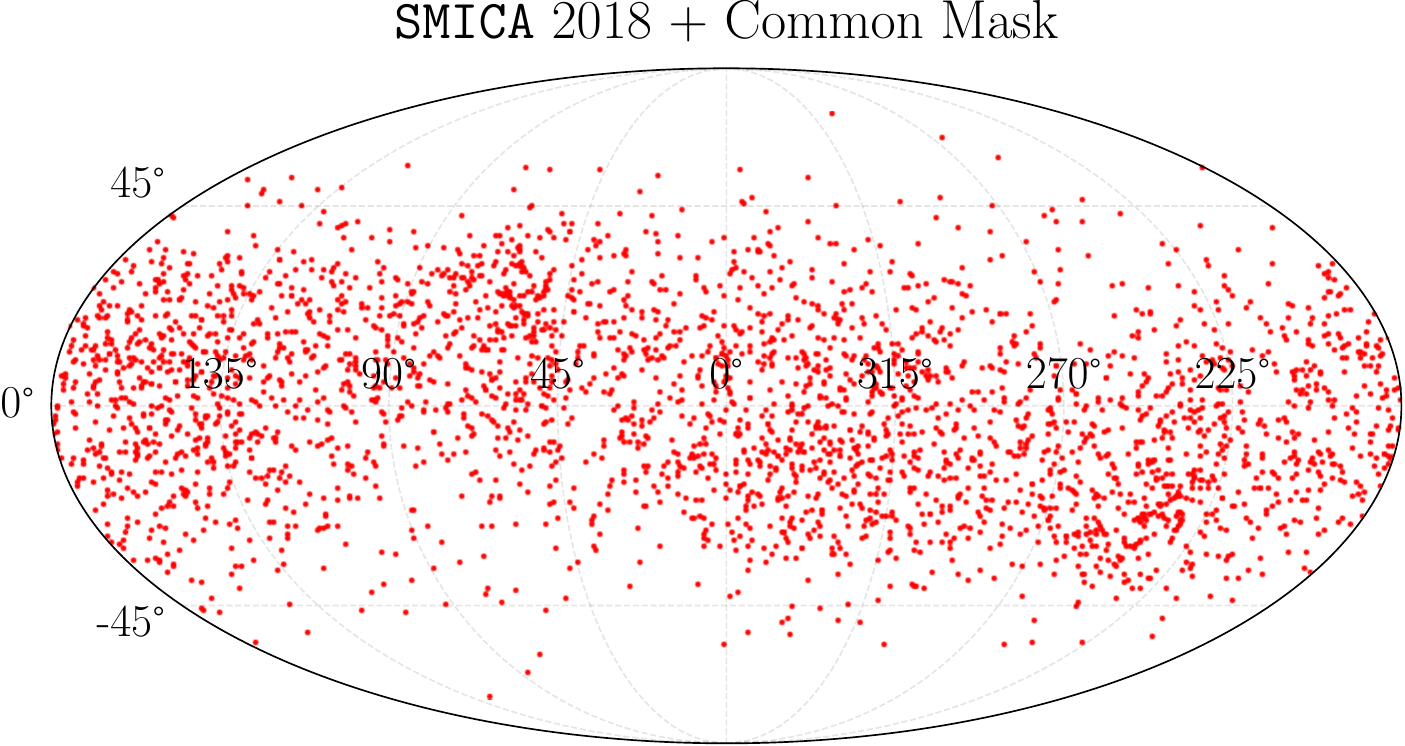}
    \caption{Similar to Figure~\ref{mollweide-mvs-fvs-fullsky} for the Planck 2018 maps after applying the Common Mask, together with a masked Gaussian and Isotropic map simulation. The inclusion of a mask leads to a small clustering of MVs on the poles (see also Figure~\ref{hist-mask}). The FVs  will then concentrate around the equator, since this is the position which minimizes their distance to the MVs.}
    \label{mollweide-mvs-fvs-masked}
\end{figure*}

\begin{figure*}[t!]
    \centering
    \includegraphics[scale=0.27]{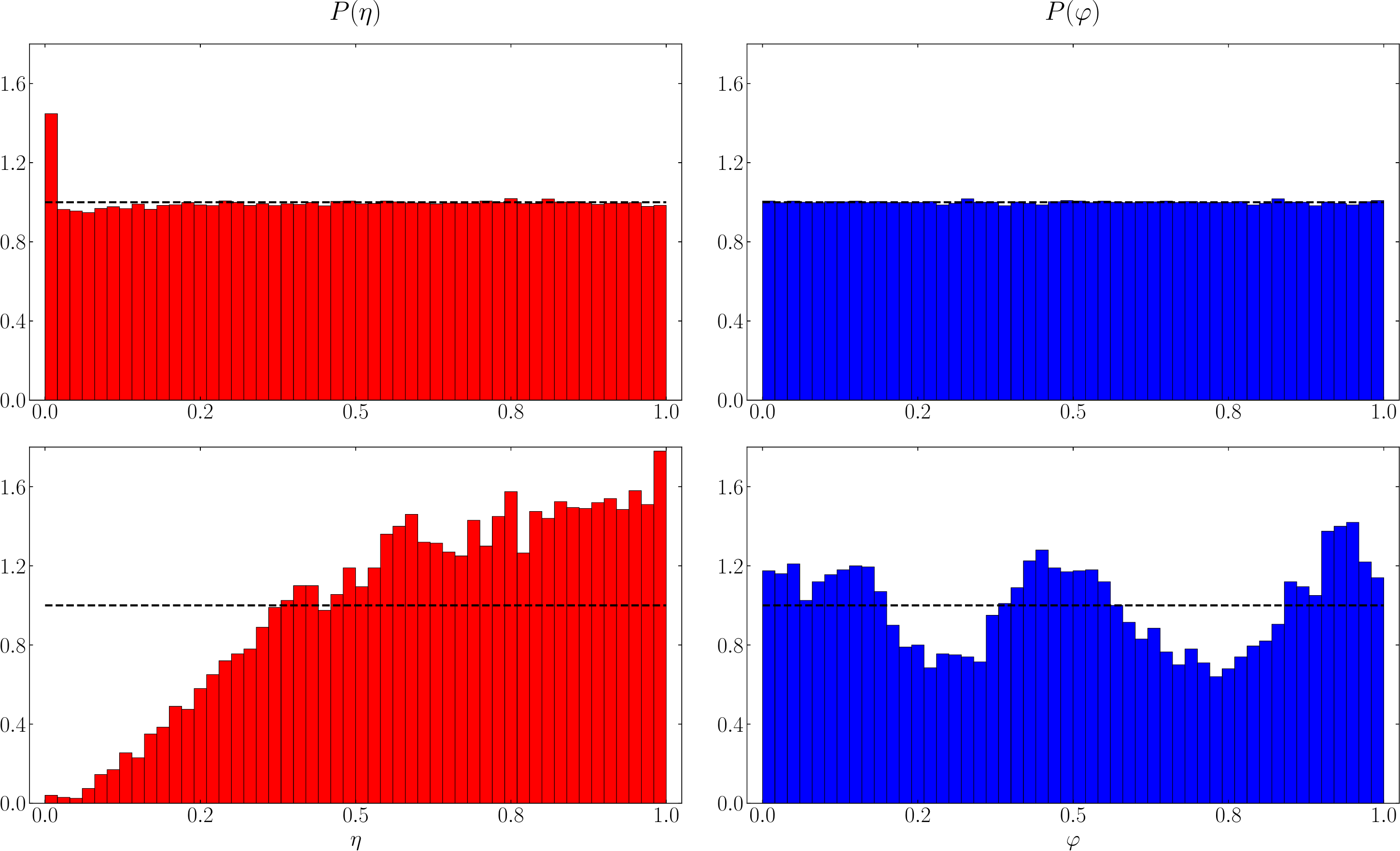}
    \caption{\emph{Top:} distribution of MV coordinates for $\ell\in[2,100]$ from 100 simulations with the 2018 Common Mask. Vectors concentrate on the north-pole ($\eta=0$) and are surrounded by a depression ring of $\sim15^\circ$ radius. \emph{Bottom:} distribution of the FVs coordinates from the corresponding MVs of the upper panel.
    \label{hist-mask}}
\end{figure*}

The inclusion of the (apodized) Planck 2018 Common Temperature Mask~\citep[henceforth Common Mask,][]{Akrami:2018mcd} makes the MVs all indistinguishable among themselves and from the isotropic simulations; this is depicted in Figure~\ref{mollweide-mvs-fvs-masked} for one isotropic simulation and two Planck maps (chosen arbitrarily), where we can see that the inclusion of a mask leads to small anisotropies near the poles (see also Figure~\ref{hist-mask}).

Note that the 2018 \texttt{Commander} MVs differs drastically from the 2015 ones. Indeed, as remarked by the Planck team, the use of full-frequency maps (as opposed to single bolometer maps) leads to a simpler foreground model employed to the \texttt{Commander} 2018 map, which includes only four different components in temperature, instead of the seven components used in the 2015 map~\citep{Akrami:2018mcd}. Figure~\ref{hist-mask} shows the effect of foreground masks on the distribution of the multipole and Fréchet vectors. Because the relation between the $a_{\ell m}$s and the MVs is non-linear, the effect of a mask on the latter is hard to predict, as confirmed by the top panel in this figure, which shows a tendency of the MVs to accumulate on the North-Pole. The FVs, on the other hand, concentrate around the equator, as can be seen in Figure~\ref{mollweide-mvs-fvs-masked}. The FV in all scales exhibit small differences between data and the isotropic simulations, which we will discuss in Section~\ref{sec:results}. Notice that the azimuthal distribution of the FVs (bottom-right panel in Figure~\ref{hist-mask}) captures the asymmetries of the Common Mask which are not obvious in the one-point distribution of the MVs, but which are nonetheless hidden in their inner correlations.

\section{Statistical Tests}

Under the null statistical hypothesis (i.e., a Gaussian, homogeneous and isotropic universe) both MVs and FVs are uniformly distributed. Thus, in order to conduct a null test of statistical isotropy, we simulate $3000$ CMB maps with $N_{side}=1024$ (see~\ref{app:nside} for more details), which are then masked with the Common Mask. After extracting the $a_{\ell m}$s in the range ${\ell\in[2,1500]}$, we obtain the MVs and their FVs as described above. We verified that, while the addition of Gaussian and \emph{isotropic} random noise to the $a_{\ell m}$s might eventually lead to drastic displacements of a few individual MVs (due to the ill-conditioning of the polynomials at some particular scales), it has no observable impact on their statistical distributions. This is expected as per the preceding discussion: such noise effectively equates to simple re-scaling of the $C_\ell$s.

The null hypothesis can then be verified by means of a simple chi-square test which, for the \emph{multipole vectors}, is described as follows: for each simulation and at each $\ell$, we obtain normalized histograms for the variables $\eta$ and $\varphi$, which give an estimate of their one-point functions. From these histograms we compute the mean number of events at the $i$-th angular bin, and the mean covariance among bins, $C_{ij}$. These quantities allow us to the define the (reduced) chi-square function:
\begin{equation}\label{chi-square}
    \chi^2_\ell(x) \!=\! \frac{1}{N_{\rm bins}(\ell)-1} \!\! \sum_{i,j}^{N_{\rm bins}-1} \! (x_i-\bar{x}_i)(C^{-1})_{ij}(x_j-\bar{x}_j),
\end{equation}
where $x_i$ stands for the number counts in the $i$-th bin of either $\eta$ or $\varphi$. Note that, because $\sum^{N_{\rm bins}}_i x_i = \ell$, not all bins are independent. We thus drop one bin\footnote{Chosen, for convenience, to be the last bin.} in the computation of each $C_{ij}$, and use $(N_{\rm bins}-1)$ as the number of independent degrees of freedom. Finally, although one can test the isotropy independently for each variable, we here focus on just the global quantity combining both:
\begin{equation}\label{chi-square-eta-plus-phi}
    \chi^2_\ell \equiv  \frac{1}{2} \big[\chi^2_\ell(\eta)+ \chi^2_\ell(\varphi)\big]\,.
\end{equation}
This can then be applied to test the uniformity of the MVs of any CMB map.

We used our 3000 simulations to determine $C_{ij}$. At this point, one would expect Eq.~\eqref{chi-square} to give $\chi^2_{\ell}\approx 1$ when applied to an independent and identically generated CMB map. This expectation is only approximately met, reflecting the fact that the overall amplitudes of the elements of $C_{ij}$ have poorly converged after these $3000$ simulations. But we confirmed that $\chi^2_\ell\rightarrow 1$ as the number of simulations increased (see~\ref{app:number-of-sims} for more details). Even though the employed algorithm is efficient, running many more simulations at high-$\ell$ is still very intensive, so we chose a more feasible solution by generating \emph{control simulations} to calibrate Eq.~\eqref{chi-square}. We have thus generated $2000$ additional (and independent) CMB maps from which we extracted all MVs, and to which we applied Eqs.~\eqref{chi-square} and~\eqref{chi-square-eta-plus-phi}. This gives us a mean theoretical $\chi^2_\ell$ \emph{at each multipole}, as well as the measure of the cosmic variance.

For the number of angular bins, $N_{\rm bins}$, a higher value allows one to better capture the fine effects of a mask (see Figure~\ref{hist-mask}) but can lead to numerical instabilities in the inversion of the covariance matrix. The most straightforward choice would be to use $N_{\rm bins} = \ell$ as there are $\ell$ multipole vectors at a given multipole $\ell$, but this leads to too much numerical noise for higher $\ell$s. We tested that for the number of simulations we used for the highest $\ell$ (1500) a total of around 600 bins gave the optimal trade-off. We thus settled on the following scheme, which keeps $N_{\rm bins} = \ell$ only until $\ell_{\rm max}=30$ (see below on the relevance of this number) and then increases linearly the number of bins until reaching 600 at $\ell=1500$:
\begin{equation}\label{Nbins}
N_{\rm bins}(\ell)=
\begin{cases}
	\ell\,,\hspace{3.42cm} \ell\leq 30 \\
	\left\lceil    \frac{57}{147}\left(\ell-30\right)+30 \right\rceil
	\,,\quad\ell> 30
\end{cases}
\end{equation}
where $\lceil \cdot \rceil$ denotes the ceiling function.

In summary, given a range of multipoles, the above algorithm will produce a list of chi-square values for the MVs (one value for each $\ell$). We then further convert this list into one number describing the overall null-hypothesis, as follows: given a list of chi-square values in a chosen range $\Delta\ell=\ell-2$, we computed the overall fit of the chi-square of the data to the chi-square of the simulations. This quantity, which we dub $\chi^2_{\rm MV}$, is given by
\begin{equation}\label{eq:chisq-chisq}
\begin{aligned}
    &\chi^2_{\rm MV}\equiv \\
    &\sum_{\ell_1,\ell_2=2}^{\ell_{\rm max}}\left[\chi^{2,\rm data}_{\ell_1} - \chi^{2, \rm sim}_{\ell_1}\right]
    C^{-1}_{\rm MV,\,\ell_1\ell_2}\left[\chi^{2,\rm data}_{\ell_2} - \chi^{2, \rm sim}_{\ell_2}\right]\!,
\end{aligned}
\end{equation}
where $C_{\rm MV}$ gives the correlation matrix between the $\chi^{2,\rm sim}_{\ell}$s at different scales. This correlation is only non-diagonal if a mask is used, as masks will in general induce correlation among different multipoles. We however verified numerically that for the Planck mask the non-diagonal terms of $C_{\rm MV}$ are negligible. We illustrate this in~\ref{app:corr-matrix}. Thus, when computing Eq.~\eqref{eq:chisq-chisq}, we can treat the multipoles as approximately independent from each other. Note that we include the parameter $\chi^{2, \rm sim}_\ell$ in order to account for the fact that $\chi^{2, \rm sim}_{\ell}\approx 1$ but it is not exactly unity, as discussed above. Due to the negligible correlations among $\ell$s  we can combine the $p$-values for each $\ell$ into a global $p$-value in a straightforward manner using Fisher's method~\citep{fisher1992statistical,brown1975400}. We then tested at different values of $\ell$ what was the probability distribution for our $\chi^2_\ell(x)$. It is in general well described by a chi-squared distribution, and in fact for small scales by a Gaussian distribution.

The uniformity of the Fréchet vectors can be tested in exactly the same manner, except that in this case Eqs.~\eqref{chi-square} and~\eqref{chi-square-eta-plus-phi} are not computed for each $\ell$, but instead for a single histogram containing all FVs in the multipolar range of interest. We thus have a \emph{a single} chi-square for all the FVs in that range:
\begin{equation}\label{chi-square-FV}
    \chi^2_{\rm FV}(x) = \frac{1}{N_{\rm bins}-1} \! \sum_{i,j}^{N_{\rm bins}-1}(x_i-\bar{x}_i)(C_{\rm FV}^{-1})_{ij}(x_j-\bar{x}_j).
\end{equation}
The FVs angular binning is performed similarly, except that in this case we replace the multipole $\ell$ in Eq.~\eqref{Nbins} by a range of multipoles. Thus, for example, to test all FVs in the range $\ell\in[2,600]$ we use $N_{\rm bins}(600-1)=251$.

While we could compute our $p$-values using these distributions, some of the unmasked maps exhibit high-levels of anisotropy, and the data points fall far in the tail of the distributions. Due to these extreme cases, in order to be very conservative when analysing the statistics of MVs we relied \emph{only} on the histograms themselves and not on the fitted distributions, and thus put a lower bound of 1/2000 on the resulting probabilities for each $\ell$. Nevertheless, since we have basically 2000 simulations for $\ell = [2,1500]$, this is still a very low lower bound. The minimum possible combined $p$-value can be computed from (using Fisher's method):
\begin{equation}\label{eq:min-p-value}
    x_{\rm Fisher} \equiv -2 \sum_{\ell = 2}^{\ell_{\rm max}} \log (1/2000) \,.
\end{equation}
The corresponding $p$-value${}_{\rm min}$ is computed as the probability of having a value of at least $x_{\rm Fisher}$ for a $\chi^2$ distribution with $2\ell_{\rm max}-2$ degrees of freedom. The result is $p$-value${}_{\rm min} \sim 10^{-2980}$. Since such tiny numbers are not very intuitive, we write all of our $p$-values in terms of the corresponding \emph{Gaussian standard deviations} $\sigma$:
\begin{equation}\label{eq:pvalues-to-sigma}
    \sigma\textrm{-value} \,\equiv\, \sqrt{2} \,{\rm Erf}^{-1} \big(1-p\textrm{-value} \big)\,.
\end{equation}
This means that our $p$-value${}_{\rm min}$ corresponds to a maximum $\sigma$-value of around 117$\sigma$. Of course if one does rely on the fitted distributions on their tails there is no limit.

For the Fréchet Vectors, since each is already a mean among all MV in a given $\ell$, their $p$-value${}_{\rm min}$ is simply 1/2000, which corresponds to a $\sigma$-value of 3.5$\sigma$. Some of our results however show a much higher discrepancy than this, so for the FV we also quote the z-score (also referred to as standard score), which is simply the number of standard deviations between the observed result and the simulated one:
\begin{equation}\label{eq:zscore}
    \text{z-score} \,\equiv\, \frac{\chi^{2, \rm data}_{\rm FV}-\chi^{2, \rm sim}_{\rm FV}}{\sigma_{\chi^2}}\,.
\end{equation}

\section{Results}\label{sec:results}

Aiming to mitigate possible \emph{a posteriori} effects we chose three well-motivated values of $\ell_{\rm max}$ even before the analysis began. We chose first $\ell_{\rm max} = 1500$, covering all-scales measured by Planck while still trying to avoid the effects of the noise anisotropy, which becomes more important at high-$\ell$. As we show below, we do not detect any anisotropies with the MV statistics, but for the FVs we notice an increasing anisotropy for $\ell \ge 1300$, a strong hint of anisotropic noise detection. In fact, for $\ell =1350$ in all 4 mapmaking pipelines we have a noise spectrum $N_{\ell} \simeq 0.1 C_\ell$~\citep{Akrami:2018mcd}. A $10\%$ contribution in a sufficiently sensitive test will result in anisotropic results.  So for the FVs we also quote results up to $\ell_{\rm max} = 1200$.  We also chose $\ell_{\rm max} = 600$ which represents the range of scales covered by WMAP; WMAP data resulted in most of the claimed anomalies still investigated today. Finally we chose $\ell_{\rm max} = 30$ to depict the results on the large-scales only. This value is the one used by the Planck team in order to separate their low-$\ell$ and high-$\ell$ likelihoods~\citep{Aghanim:2015xee}.

\begin{figure}[t!]
    \centering
    \includegraphics[width = \columnwidth] {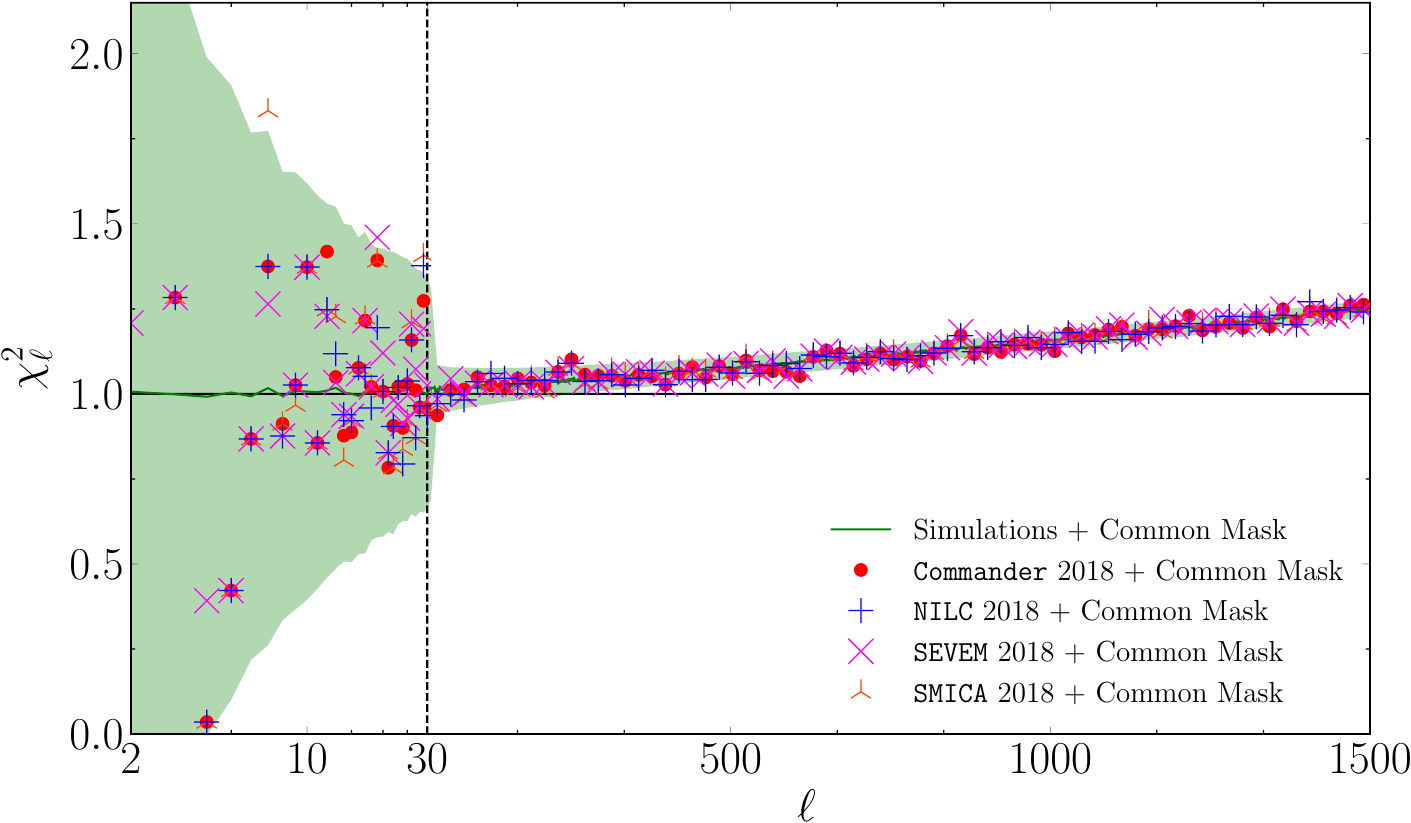}
    \caption{Test of isotropy as a function of $\ell$ for the 4 masked Planck 2018 maps. The solid (green) curve gives $\chi_\ell^2$  averaged over $2000$ control simulations, and the green bands show $2\sigma$ cosmic variance. For clarity, data points are gathered in 49 bins of $\Delta\ell=30$ in the interval ${\ell\in[31,1500]}$.}
	\label{masked-results}
	\medskip
    \footnotesize
    \setlength\tabcolsep{4pt}
    \begin{tabular}{cccccc}
        \toprule
        \multicolumn{2}{c}{\textbf{Masked MVs (PR3)}} & \!\!\texttt{Commander}\!\! & \texttt{NILC} & \texttt{SEVEM} & \texttt{SMICA} \\
        \midrule
        \midrule
        \multirow{2.3}{*}{Large scales} & $\chi^2_{\rm MV}/\text{dof}$ & 0.811 & 0.765 & 0.579 & 1.12 \\
        \cmidrule{2-6}
        & $\sigma\text{-value}$ & \textbf{0.20} & \textbf{0.14} & \textbf{0.02} & \textbf{0.86} \\
        \midrule
        \multirow{2.3}{*}{WMAP scales} & $\chi^2_{\rm MV}/\text{dof}$ & 1.06 & 0.904 & 1.04 & 1.04 \\
        \cmidrule{2-6}
        & $\sigma\text{-value}$ & \textbf{1.2} & \textbf{0.05} & \textbf{1.1} & \textbf{1.3}\\
        \midrule
        \multirow{2.3}{*}{All scales} & $\chi^2_{\rm MV}/\text{dof}$ & 0.998 & 0.977 & 1.03 & 0.968 \\
        \cmidrule{2-6}
        & $\sigma\text{-value}$ & \textbf{0.54} & \textbf{0.24} & \textbf{1.1} & \textbf{0.27}\\
        \bottomrule
    \end{tabular}
        \captionof{table}{Goodness-of-fit of the data points of Figure~\ref{masked-results} and their associated $\sigma$-values (${\rm d.o.f.}=\ell_{\rm max} -1$). The analysis was divided in three ranges of interest: Large scales ($\ell\in[2,30]$), WMAP scales ($\ell\in[2,600]$) and ``All'' scales ($\ell\in[2,1500]$). All 4 pipelines appear isotropic under this test.}
    \label{table-chi2-masked}
\end{figure}

\begin{figure}[t!]
    \centering
    \includegraphics[width = \columnwidth] {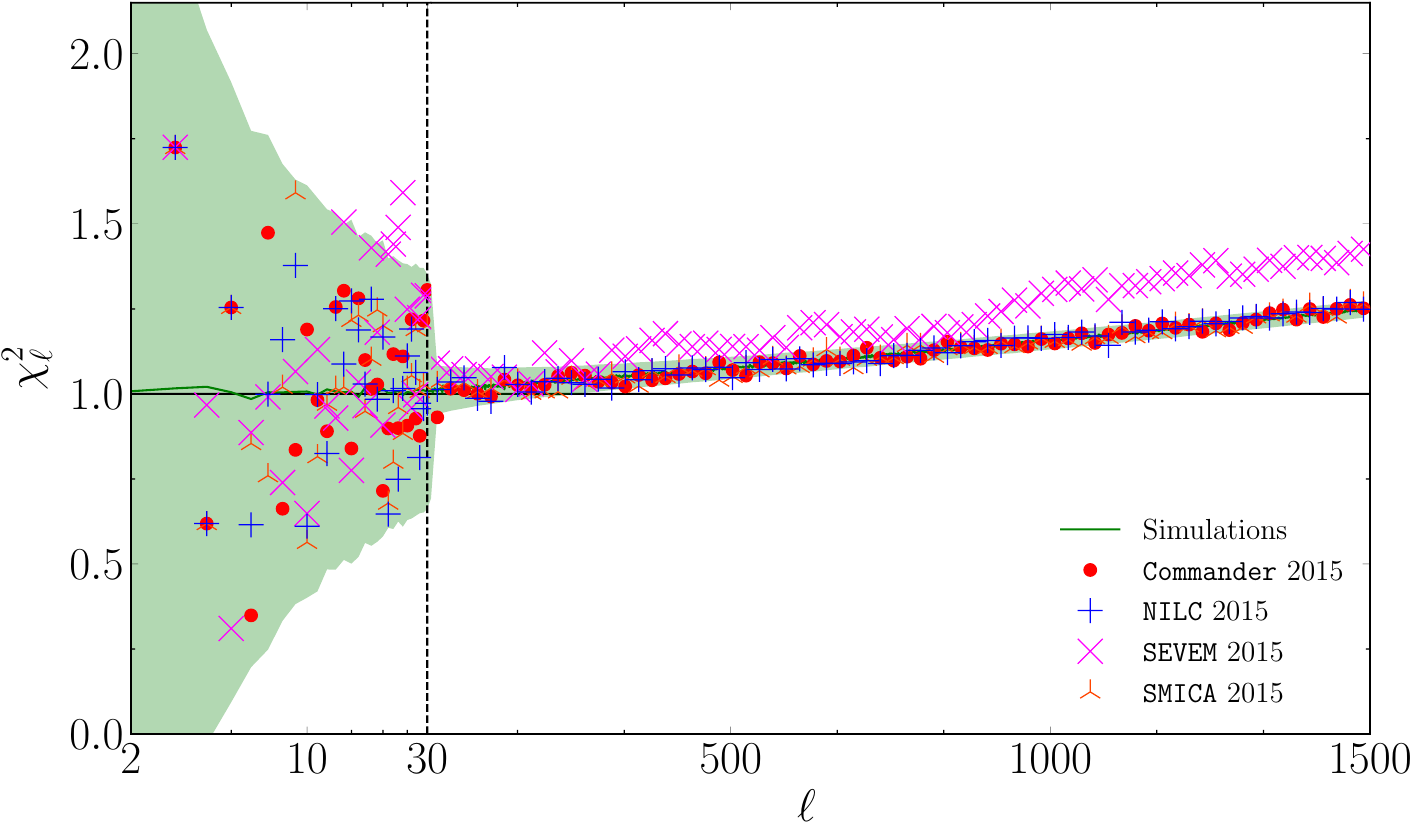}
    \caption{Same as Figure~\ref{masked-results} but for the unmasked, full sky Planck 2015 maps. The full sky \texttt{SEVEM} map is in disagreement with the isotropy hypothesis at over $47\sigma$. See Table~\ref{table-chi2-fullsky-2015}.}
	\label{fullsky-2015-results}
	\medskip
    \footnotesize
    \setlength\tabcolsep{4pt}
    \begin{tabular}{cccccc}
        \toprule
        \multicolumn{2}{c}{\textbf{Full sky MVs (PR2)}} & \!\!\texttt{Commander}\!\! & \texttt{NILC} & \texttt{SEVEM} & \texttt{SMICA}\\
        \midrule
        \midrule
        \multirow{2.3}{*}{Large scales} & $\chi^2_{\rm MV}/\text{dof}$ & 0.933 & 0.807 & 1.83 & 0.808 \\
        \cmidrule{2-6} & $\sigma\text{-value}$ & \textbf{0.58} & \textbf{0.25} & \textbf{2.6} & \textbf{0.20}\\
        \midrule
        \multirow{2.3}{*}{WMAP scales} & $\chi^2_{\rm MV}/\text{dof}$ & 1.01 & 0.907 & 1.60 & 0.956\\
        \cmidrule{2-6}
        & $\sigma\text{-value}$ & \textbf{0.63} & \textbf{0.10} & \textbf{8.4} & \textbf{0.25}\\
        \midrule
        \multirow{2.3}{*}{All scales} & $\chi^2_{\rm MV}/\text{dof}$ & 0.983 & 1.00 & 4.26 & 1.08\\
        \cmidrule{2-6}
        & $\sigma\text{-value}$ & \textbf{0.36} & \textbf{0.74} & $\boldsymbol{>}$\textbf{47} & \textbf{2.2}\\
        \bottomrule
    \end{tabular}
    \captionof{table}{Same as Table~\ref{table-chi2-masked} but for the unmasked Planck 2015 data. Only the \texttt{SEVEM} map shows deviations from isotropy (already at WMAP scales) in this test. See the text for more details.
	\label{table-chi2-fullsky-2015}}
\end{figure}

\begin{figure}[t!]
	\centering
    \includegraphics[width = \columnwidth] {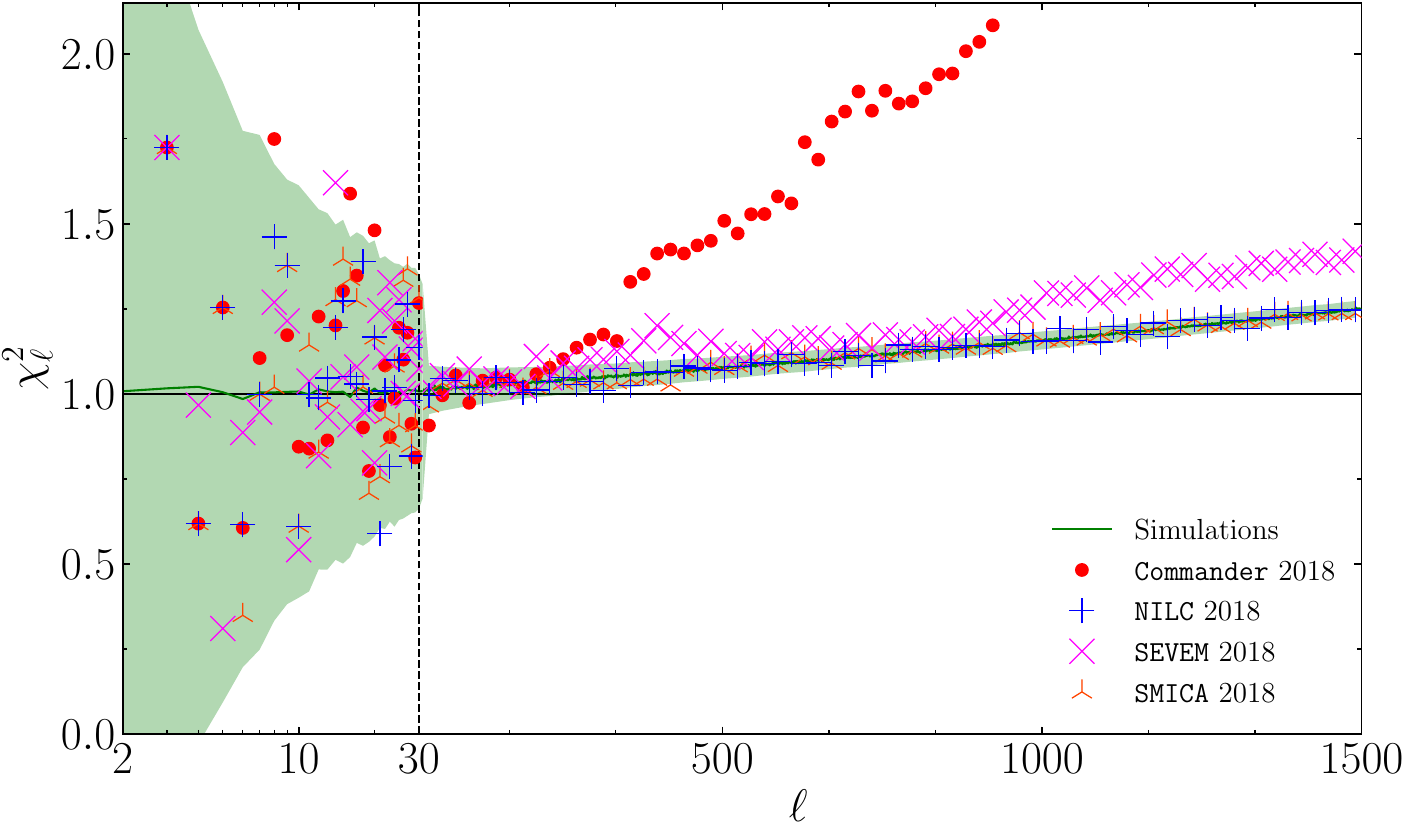}
    \caption{Same as Figure~\ref{masked-results} but for unmasked Planck 2018 maps. The full sky \texttt{SEVEM} and \texttt{Commander} maps are in disagreement with the isotropy hypothesis already at the WMAP scales. At All scales, the discrepancy is over $99\sigma$ and $40\sigma$, respectively. See Table~\ref{table-chi2-fullsky-2018}.}
	\label{fullsky-2018-results}
	\medskip
    \footnotesize
    \setlength\tabcolsep{4pt}
    \begin{tabular}{cccccc}
        \toprule
        \multicolumn{2}{c}{\textbf{Full sky MVs (PR3)}} & \!\!\texttt{Commander}\!\! & \texttt{NILC} & \texttt{SEVEM} & \texttt{SMICA}\\
        \midrule
        \midrule
        \multirow{2.3}{*}{Large scales} & $\chi^2_{\rm MV}/\text{dof}$ & 1.35 & 0.962 & 0.985 & 1.27\\
        \cmidrule{2-6} & $\sigma\text{-value}$ & \textbf{1.4} & \textbf{0.49} & \textbf{0.52} & \textbf{1.3}\\
        \midrule
        \multirow{2.3}{*}{WMAP scales} & $\chi^2_{\rm MV}/\text{dof}$ & 12.6 & 0.973 & 1.39 & 1.01 \\
        \cmidrule{2-6}
        & $\sigma\text{-value}$ & $\boldsymbol{>}$\textbf{42} & \textbf{0.47} & \textbf{5.7} & \textbf{0.66}\\
        \midrule
        \multirow{2.3}{*}{All scales} & $\chi^2_{\rm MV}/\text{dof}$ & 215 & 1.02 & 3.63 & 1.01 \\
        \cmidrule{2-6}
        & $\sigma\text{-value}$ & $\boldsymbol{>}$\textbf{99} & \textbf{1.2} & $\boldsymbol{>}$\textbf{40} & \textbf{0.74}\\
        \bottomrule
	\end{tabular}
    \captionof{table}{Same as Table~\ref{table-chi2-masked} but for the unmasked Planck 2018 data. Compared to 2015, \texttt{SEVEM} shows a slight improvement, but \texttt{Commander} becomes completely anisotropic. 	\label{table-chi2-fullsky-2018}}
\end{figure}

\subsection{Multipole Vectors}

Figure~\ref{masked-results} summarizes the result of our analysis applied to the four 2018 masked Planck maps. We stress that these results are totally independent of existing measurements of the $C_{\ell}$s, and thus of any of its claimed anomalies, as well as of the addition of isotropic instrumental noise, regardless of the amplitude of its spectrum. Table~\ref{table-chi2-masked} gives the global goodness-of-fit of the data points in the Figure~\ref{masked-results} with respect to the theoretical curve. Under this test, all four masked Planck pipelines are consistent with isotropy. In particular, no deviation of isotropy was detected at large scales ($\ell\in[2,30]$), where most of CMB anomalies were reported.

The Planck collaboration provides a Common Mask for the community to allow the use of their data safe from foreground contamination and advises one not to use their maps without this mask. However, here testing the isotropy of the full sky Planck maps is interesting because it provides a good test of the power of the method to detect these foregrounds. After all, these regions have allegedly high levels of foreground contamination.

We thus present the results of the same analysis constructed with full sky simulations and applied to the four full sky Planck pipelines. In Figure~\ref{fullsky-2015-results} and Table~\ref{table-chi2-fullsky-2015} we show the results for the full sky 2015 maps, and in Figure~\ref{fullsky-2018-results} and Table~\ref{table-chi2-fullsky-2018} the same for 2018. Interestingly, all full sky maps (except SEVEM 2015, which shows a moderate deviation) are consistent with the isotropy hypothesis at large scales ($\ell\in[2,30]$), where most of known CMB anomalies were reported. This is not a counter-proof of existing $C_\ell$ anomalies since, again, our results are independent of this quantity. We find that 2015 \texttt{SEVEM} and 2018 \texttt{SEVEM} and \texttt{Commander} maps are in flagrant disagreement with the isotropy hypothesis, becoming highly anisotropic at $\ell\gtrsim300$. The differences between the two releases of \texttt{Commander} can be attributed to the simpler foreground model employed by the Planck team in the 2018 release~\citep{Akrami:2018mcd}. Regarding \texttt{SEVEM}, both full sky releases show strong traces of residual contaminations, although the 2018 release shows a slight improvement ($\chi^2_{\rm MV}/{\rm d.o.f}\approx 3.6$) in comparison to the 2015 release ($\chi^2_{\rm MV}/{\rm d.o.f}\approx 4.3$). These findings are in agreement with the recent results of~\citet{Pinkwart:2018nkc,Minkov:2018prz}, where \texttt{Commander} (2018) and \texttt{SEVEM} (2015 and 2018) maps were also found to contain anisotropic residuals.

Overall, our analysis with full sky maps, which are known to contain anisotropic residuals, show no deviations of isotropy in both releases of \texttt{NILC} and \texttt{SMICA}, and the 2015 release of \texttt{Commander}.
This should not be seen as a weakness of the MVs to detect anisotropies on these maps, but rather as a limitation of the simple chi-square test when applied to the one-point distribution of these vectors, which ignores the correlations between the MVs at a given multipole. As we show below, the same analysis conducted with the Fréchet vectors clearly pinpoints the presence of anisotropies in all full sky maps at small scales. On the other hand, the fact that \texttt{NILC} and \texttt{SMICA} appear isotropic even without masks suggests that, for some applications, one may rely on smaller masks than Planck's Common Mask. We investigate this possibility in more detail in~\ref{app:inpainted}, where we apply instead of this Common Mask the much smaller inpainting mask, which removes only $2.1\%$ of the sky, and show that this is sufficient to make all 4 pipelines in agreement with the isotropic hypothesis in our MV statistic.

\subsection{Fréchet Vectors}

We recall that, since we have only one Fréchet vector at each $\ell$ (as opposed to $\ell$ MVs per $\ell$), we gather all vectors from a multipolar range
$[\ell_{\rm min},\,\ell_{\rm max}]$ into a single histogram. When compared to the theoretical (i.e., simulated) distribution of Fréchet vectors, this leads to one chi-square value per multipole range (instead of one chi-square value per single $\ell$, as in Figures~\ref{masked-results}--\ref{fullsky-2018-results}).

\begin{figure}[t!]
	\centering
    \includegraphics[width = \columnwidth]{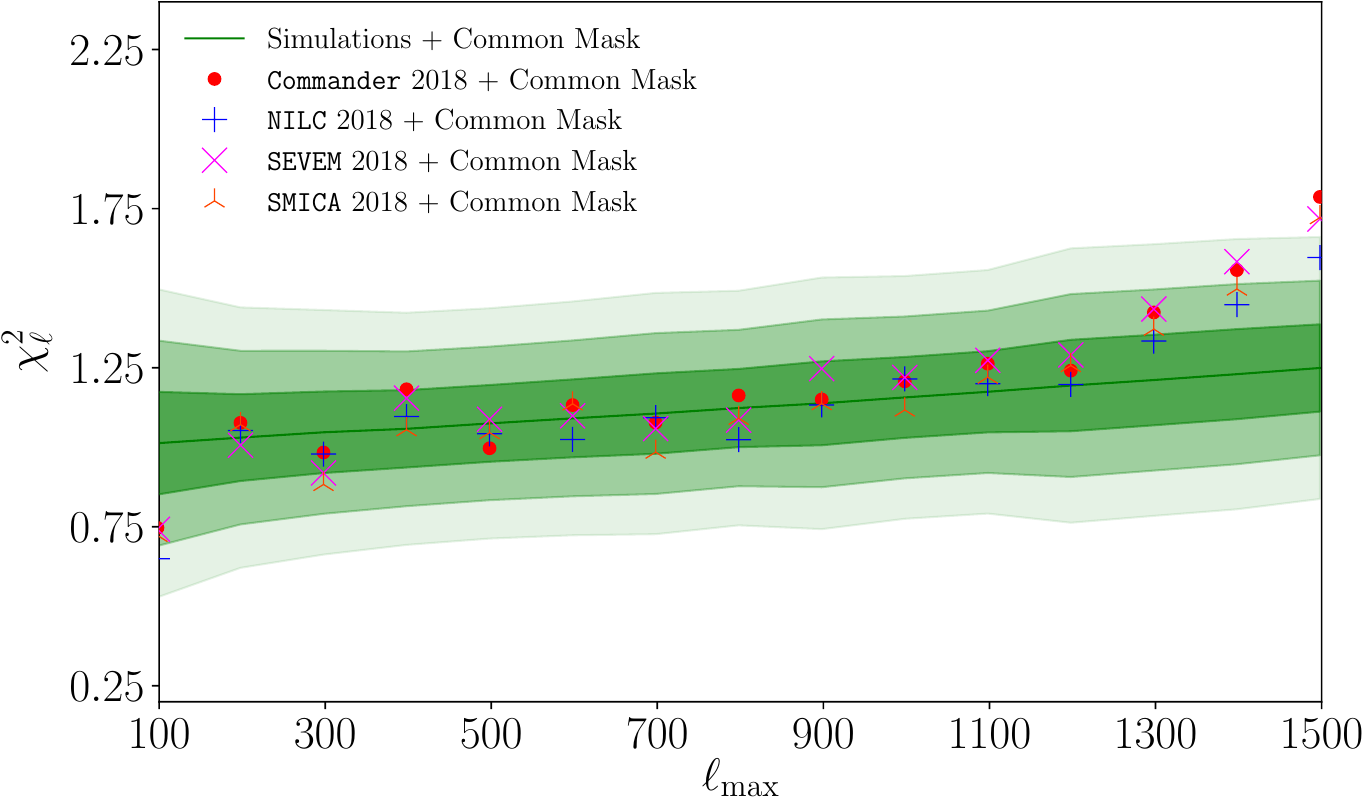}
    \caption{Test of isotropy for the Fréchet vectors of masked Planck maps as a function of $\ell_{\rm max}$. Green bands stand for 1, 2 and 3$\sigma$ cosmic variance regions. For $\ell > 1200$ anisotropies start to significantly increase, likely due to the anisotropy in the noise which become more important at higher $\ell$.}
	\label{masked-frechet-plot}
    \medskip
    \renewcommand{\arraystretch}{1.45}
    \footnotesize
    \setlength\tabcolsep{4pt}
    \begin{tabular}{cccccc}
        \specialrule{.8pt}{2pt}{0pt}
        \multicolumn{2}{c}{\textbf{Masked FVs (PR3)}} & \!\!\texttt{Commander}\!\! & \texttt{NILC} & \texttt{SEVEM} & \texttt{SMICA} \\
        \specialrule{.5pt}{2pt}{0pt}
        \specialrule{.5pt}{2pt}{0pt}
        \multirow{2}{*}{Large scales} & $\chi^2_{\rm FV}/\text{dof}$ & 1.15 & 1.16 & 1.07 & 1.22 \\
        \cline{2-6}  & $\sigma$-value & \textbf{1.1} & \textbf{1.1} & \textbf{0.84} & \textbf{1.3} \\
        \hline
        \multirow{2}{*}{WMAP scales} & $\chi^2_{\rm FV}/\text{dof}$ & 1.13 & 1.02 & 1.10 & 1.14 \\
        \cline{2-6} & $\sigma$-value & \textbf{0.90} & \textbf{0.38} & \textbf{0.72} & \textbf{0.93}\\
        \hline
        \multirow{2}{*}{$\ell\in[2,1200]$} & $\chi^2_{\rm FV}/\text{dof}$ & 1.24 & 1.20 & 1.29 & 1.26 \\
        \cline{2-6}  & $\sigma$-value & \textbf{0.89} & \textbf{0.69} & \textbf{1.1} & \textbf{0.97}\\
        \hline
        \multirow{2}{*}{All scales} & $\chi^2_{\rm FV}/\text{dof}$ & 1.79 & 1.60 & 1.72 & 1.72\\
        \cline{2-6} & $\sigma$-value & $\boldsymbol{>}$\textbf{3.5} & \textbf{2.8} & \textbf{3.5} & \textbf{3.5}\\
        \hline
    \end{tabular}
    \captionof{table}{Same as Table~\ref{table-chi2-masked} but for the  Fréchet vectors. The FV are more sensitive to anisotropies, and for scales $\ell\geq 1300$ it starts to detect anisotropies in all maps, likely due to the noise anisotropy. At $\ell \sim 1350$ the noise $N_\ell$ is $\sim 0.1 C_\ell$, and it grows quickly with increasing $\ell$. We therefore include here also the results for $\ell\in[2,1200]$. All maps are consistent with isotropy (at $< 2\sigma$) at scales $\ell\leq 1300$. \label{masked-frechet-results}}
\end{figure}

As highlighted in the beginning of this Section, the $\chi^2$ analysis on the FVs is more sensitive than the one on the bare MVs. The reason for this is that it takes into account the inner correlation of the MVs at each given $\ell$. Its higher sensitivity picks up an increasing anisotropic signature for $\ell>1200$ for all 4 mapmaking methods. This is illustrated in Figure~\ref{masked-frechet-plot}, where we also show the mean theoretical values together with their 1, 2 and 3$\sigma$ bands in our 2000 control simulations. At $\ell\sim 1500$ all maps are inconsistent with isotropy at around $3\sigma$. A thorough modelling of anisotropic instrumental noise is needed to confirm the origin of these anisotropies and to extend our results to higher $\ell$s. This is left for a future publication. Table~\ref{masked-frechet-results} shows the corresponding significance levels in the 3 ranges of scales considered here. Our analysis confirms the concordance of masked Planck maps with the null hypothesis at all scales $\ell\lesssim 1300$.

\begin{figure}[t!]
\centering
	\medskip
    \footnotesize
    \setlength\tabcolsep{4pt}
    \begin{tabular}{cccccc}
        \toprule
        \multicolumn{2}{c}{\textbf{Full sky FVs (PR2)}} & \!\!\texttt{Commander}\!\! & \texttt{NILC} & \texttt{SEVEM} & \texttt{SMICA}\\
        \midrule
        \midrule
        \multirow{4}{*}{Large scales} & $\chi^2_{\rm FV}/\text{dof}$ & 1.25 & 1.08 & 3.07 & 0.971\\
        \cmidrule{2-6} & $\sigma\text{-value}$ & \textbf{1.4} & \textbf{0.85} & $\boldsymbol{>}$\textbf{3.5} & \textbf{0.54}\\
        \cmidrule{2-6}
        & z-score & \textbf{1.0} & \textbf{0.3} & \textbf{9.2} & \textbf{-0.2}\\
        \midrule
        \multirow{4}{*}{WMAP scales} & $\chi^2_{\rm FV}/\text{dof}$ & 1.13 & 1.22 & 26.3 & 1.14 \\
        \cmidrule{2-6}
        & $\sigma$-value & \textbf{1.0} & \textbf{2.1} & $\boldsymbol{>}$\textbf{3.5} & \textbf{1.2}\\
        \cmidrule{2-6}
        & z-score & \textbf{0.53} & \textbf{1.8} & \textbf{350} & \textbf{0.71}\\
        \midrule
        \multirow{4}{*}{All scales} & $\chi^2_{\rm FV}/\text{dof}$ & 1.66 & 3.15 & 84.3 & 1.83 \\
        \cmidrule{2-6}
        & $\sigma$-value &$\boldsymbol{>}$\textbf{3.5} &$\boldsymbol{>}$\textbf{3.5} & $\boldsymbol{>}$\textbf{3.5} & $\boldsymbol{>}$\textbf{3.5}\\
        \cmidrule{2-6}
        & z-score & \textbf{6.9} & \textbf{33} & \textbf{1400} & \textbf{9.9}\\
        \bottomrule
	\end{tabular}
    \captionof{table}{Same as Table~\ref{masked-frechet-results} bur for the unmasked 2015 Planck maps. We also include the z-scores for WMAP scales and All Scales, which show the huge significance of the detected anisotropies in some cases. \texttt{SEVEM} is the only method which exhibits strong anisotropies also in Large scales and WMAP scales. For all scales all maps become anisotropic, but Commander and SMICA much less so.
    \label{fullsky-2015-frechet-results}}
\end{figure}

The higher sensitivity of the Fréchet vector statistic can also be seen in the unmasked, full sky maps. In Table~\ref{fullsky-2015-frechet-results} and \ref{fullsky-2018-frechet-results} we show the results for the 2015 and 2018 Planck releases, respectively. For Planck 2015, all maps are highly anisotropic when considering all scales, but the amount of anisotropy differ significantly among the different mapmaking procedures. \texttt{Commander} performs the best, with a z-score of 6.9, followed by \texttt{SMICA} and then \texttt{NILC}. \texttt{SEVEM} has a huge z-score of more than a 1000, and in particular shows anisotropies in all multipole ranges considered. For the 2018 release, some pipelines perform better, others worse. As in the MV statistic, \texttt{Commander} shows huge levels of anisotropies at WMAP scales and all scales. \texttt{SEVEM} performs only slightly better than in 2015 and \texttt{NILC} slightly worse. \texttt{SMICA} improves significantly and on all scales shows a relatively small z-score of 4.8, half the amount than in 2015.

\begin{figure}[t!]
\centering
	\medskip
    \footnotesize
    \setlength\tabcolsep{4pt}
    \begin{tabular}{cccccc}
        \toprule
        \multicolumn{2}{c}{\textbf{Full sky FVs (PR3)}} & \!\!\texttt{Commander}\!\! & \texttt{NILC} & \texttt{SEVEM} & \texttt{SMICA}\\
        \midrule
        \midrule
        \multirow{4}{*}{Large scales} & $\chi^2_{\rm FV}/\text{dof}$ & 1.53 & 1.10 & 2.18 & 1.12\\
        \cmidrule{2-6} & $\sigma$-value & \textbf{2.5} & \textbf{0.91} & $\boldsymbol{>}$\textbf{3.5} & \textbf{0.97}\\
        \cmidrule{2-6}
        & z-score & \textbf{2.3} & \textbf{0.3} & \textbf{5.2} & \textbf{0.4}\\
        \midrule
        \multirow{4}{*}{WMAP scales} & $\chi^2_{\rm FV}/\text{dof}$ & 32.4 & 1.24 & 18.3 & 1.19 \\
        \cmidrule{2-6}
        & $\sigma$-value & $\boldsymbol{>}$\textbf{3.5} & \textbf{2.4} & $\boldsymbol{>}$\textbf{3.5} & \textbf{1.8}\\
        \cmidrule{2-6}
        & z-score & \textbf{440} & \textbf{2.2} & \textbf{240} & \textbf{1.4}\\
        \midrule
        \multirow{4}{*}{All scales} & $\chi^2_{\rm FV}/\text{dof}$ & 161 & 3.55 & 60.7 & 1.53 \\
        \cmidrule{2-6}
        & $\sigma$-value & $\boldsymbol{>}$\textbf{3.5} & $\boldsymbol{>}$\textbf{3.5} & $\boldsymbol{>}$\textbf{3.5} & $\boldsymbol{>}$\textbf{3.5}\\
        \cmidrule{2-6}
        & z-score & \textbf{2700} & \textbf{39} & \textbf{1000} & \textbf{4.8}\\
        \bottomrule
	\end{tabular}
    \captionof{table}{Same as Table~\ref{fullsky-2015-frechet-results} but for the unmasked 2018 Planck maps. Notice that at WMAP scales both SEVEM and Commander exhibit strong anisotropies. Including all scales all maps become anisotropic, but SMICA considerably less so than the others.
    \label{fullsky-2018-frechet-results}}
\end{figure}

\section{Conclusions}

Our findings illustrate the usefulness of MV analyses. They constitute a great blind tool in the detection of residual anisotropic contaminations in CMB maps, which are an indication of residual foregrounds. Therefore it is possible that MV analysis can be explored to help determine which regions to mask. In order to achieve this one should ideally find a method to correlate the positions of the MVs to positions of anisotropic sources in configuration space. The novel Fréchet vectors introduced here partially accomplish this task. Indeed, Figures~\ref{mollweide-mvs-fvs-masked} show that the Fréchet vectors of masked maps fall precisely on the equatorial line where most of the mask is found.

In any case, a simpler method would be just to use MV and FV isotropy to validate and calibrate other methods of mask determination. The full sky map analyses we performed allow a quick glimpse on whether the Common Mask is too conservative. The results indicate \texttt{SMICA} is the less sensitive procedure to foregrounds at low latitudes as even in the full unmasked sky their FV had a comparatively low amount of anisotropy.

Overall, we confirm the consistency between Planck data and the fundamental hypothesis of the standard cosmological model. Our results using the bare MV in all masked Planck maps are in agreement with a isotropic and Gaussian CMB in all range of scales considered. The results using the FVs also show no anisotropy on large or WMAP scales. The absence of large-scale anisotropies is likely a consequence of our explicit avoidance of \emph{a posteriori} statistics. For $\ell\in[2,1500]$ the FV instead show an increasing anisotropy for $\ell\ge1300$, but this behavior is indicative that our statistic is detecting the well-known anisotropy in the noise. For $\ell \leq 1300$ even the more sensitive FV statistic is consistent with the isotropic and Gaussian hypothesis. In future work we plan to extend our analysis to higher $\ell$ by carefully taking into account the anisotropy in the noise in each one of the 4 mapmaking pipelines.

We stress that the main novelty of our results are not the detection of anisotropy in full sky maps (which are known to contain residual anisotropies) but the power of the formalism to detect such anisotropies at high statistical levels. Moreover, since they can also be directly applied to polarization maps, all the applications discussed here can be extended directly to all primordial CMB maps. Further research is needed to investigate this possibility in detail.

Clearly our results do not rule out all types of anisotropies. In fact, in this first analysis we are ignoring anisotropies that result on correlations between different $\ell$s, and perform only  1-point  statistical tests on the MVs and FVs (although the FVs themselves take into account the MV correlations in a given $\ell$). We leave an analysis of 2-point (or higher) statistics for future work. These correlations are common to a broad class of models, including astrophysical sources of anisotropies, aberration of the CMB and/or non-Gaussianities like lensing. Instead, our main focus was to build a simple and well-motivated statistic as much as possible free of \emph{a posteriori} selection effects. There was no guarantee that the data would pass this test, and the fact that it did favors the standard isotropic model. It also constitutes a new test that all anisotropic models must pass henceforth.\\

\section*{CRediT authorship contribution statement}
\textbf{Renan A. Oliveira:} Software, Formal analysis, Investigation, Visualization. 
\textbf{Thiago S. Pereira:} Formal analysis, Methodology, Validation, Writing - review \& editing. 
\textbf{Miguel Quartin:} Conceptualization, Methodology, Validation, Writing - review \& editing.

\section*{Acknowledgments}
	We thank Jeffrey Weeks for providing a copy of his code, Alessio Notari for important early discussions, Camila Novaes, Marvin Pinkwart, Omar Roldan and Dominik Schwarz for useful correspondence, and Marcello Oliveira da Costa for the sharing of computational resources. RAO thanks Coordenação de Aperfeiçoamento de Pessoal de Nível Superior (CAPES). TSP thanks Brazilian Funding agencies CNPq and Fundação Araucária. MQ is supported by the Brazilian research agencies CNPq and FAPERJ. This work made use of the CHE cluster, managed and funded by the COSMO/CBPF/MCTI, with financial support from FINEP and FAPERJ, and operating at Javier Magnin Computing Center/CBPF. The results of this work have been derived using HEALPix~\citep{Gorski:2004by} and Healpy packages.

\bibliographystyle{elsarticle-harv}
\bibliography{mvectors}
	
\appendix
	
\section{Computational complexity comparison}\label{app:timings}
	
\begin{figure}[t!]
    \centering
    \includegraphics[width = .95 \columnwidth]{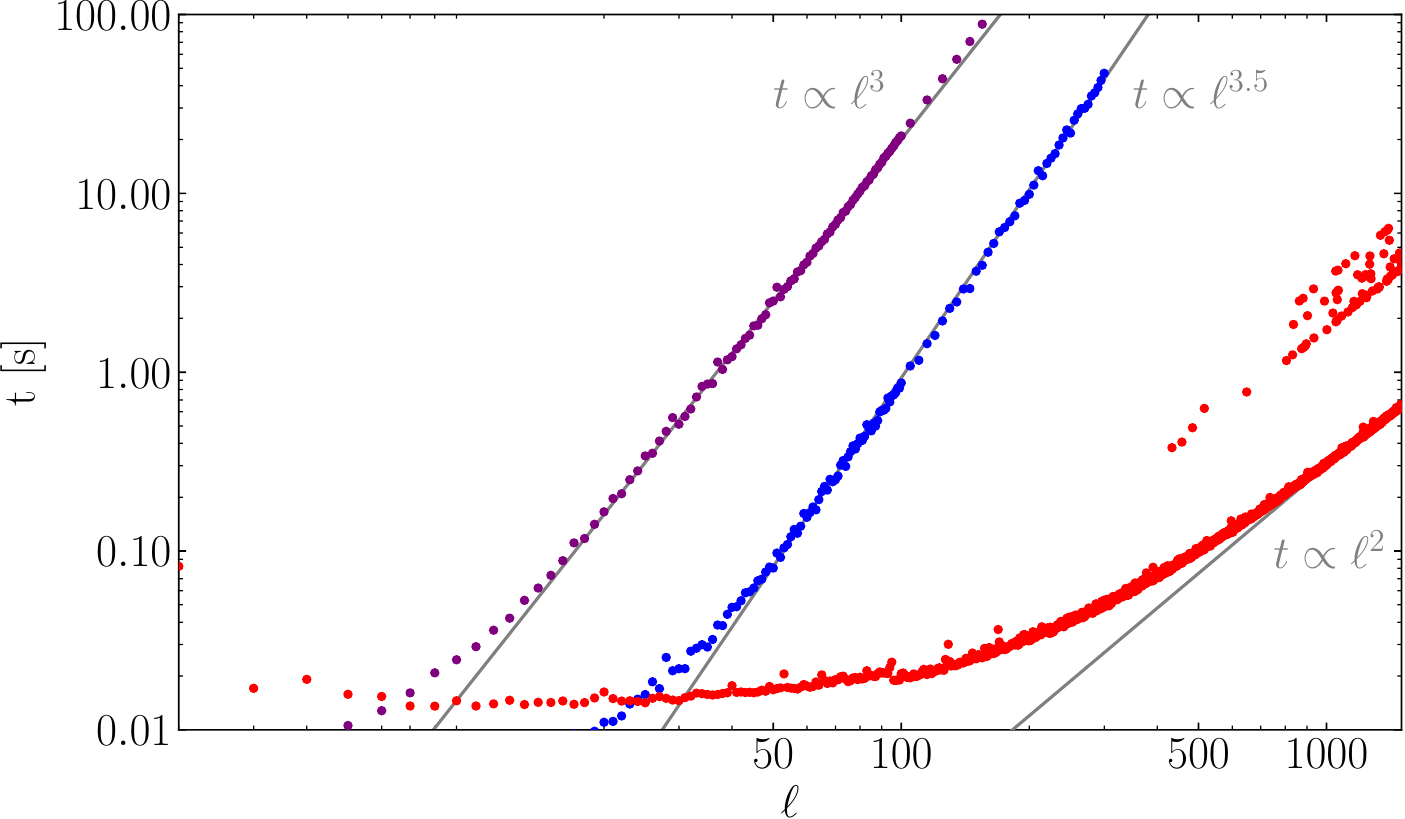}
    \caption{Computational complexity comparison. Computational time of MVs extraction using {\tt polyMV} (red) and the algorithms~\citep{Copi:2003kt} (blue) and~\citep{Weeks:2004cz} (purple). {\tt polyMV} has computational complexity ${\cal O}(\ell^2)$ as compared to ${\cal O}(\ell^{3.5})$ of both the other available codes. The scatter seen at ${\ell\gtrsim 400}$ is a consequence of numerical ill-conditioning of the polynomials at high multipoles, which demands more CPU time of {\tt MPSolve}.    \label{fig:timings}}
\end{figure}

As discussed in the main text, in this work we developed a new code {\tt polyMV} to efficiently compute all the MVs of a given map. This code is order of magnitudes faster than both existing public algorithms at high multipoles. In fact, it has computational complexity ${\cal O}(\ell^2)$ as compared to ${\cal O}(\ell^{3.5})$ of both the other available codes. Figure~\ref{fig:timings} depicts the computation time of {\tt polyMV} as a function of $\ell$ for our code as well as the other two public codes. Starting at $\ell \sim 400$ some of the polynomial of equation~\eqref{Q-poly} become ill-conditioned due to their very high-order. This is reflected on an order of magnitude longer computational time in order to evaluate their roots with the same numerical accuracy. But even in these cases the evaluation at, say, $\ell =1000$ takes less than $1\%$ of the time than the best competing code.
	
\section{Discussion on the total number of simulations}\label{app:number-of-sims}
	
\begin{figure}[t!]
    \centering
    \includegraphics[width = .95 \columnwidth] {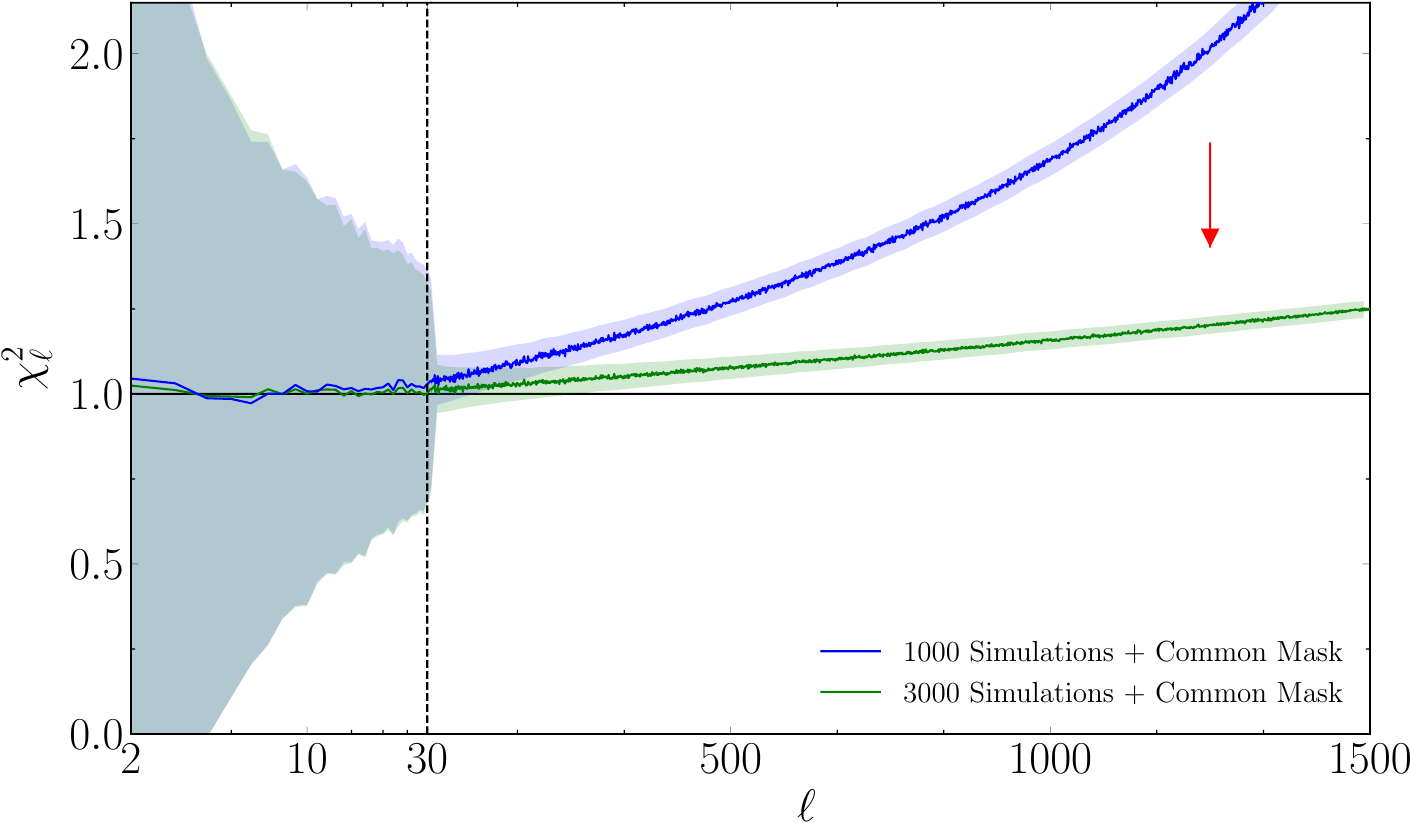}
    \caption{Comparison of the simulated Gaussian and isotropic $\chi^2_\ell$ using either 1000 and 3000 masked simulations to compute the covariance matrix $C_{ij}$ of Eq.~\eqref{chi-square}.
    	\label{fig:number-of-sims-mask}
	}
\end{figure}

\begin{figure}[t!]
    \centering
    \includegraphics[width = .95\columnwidth] {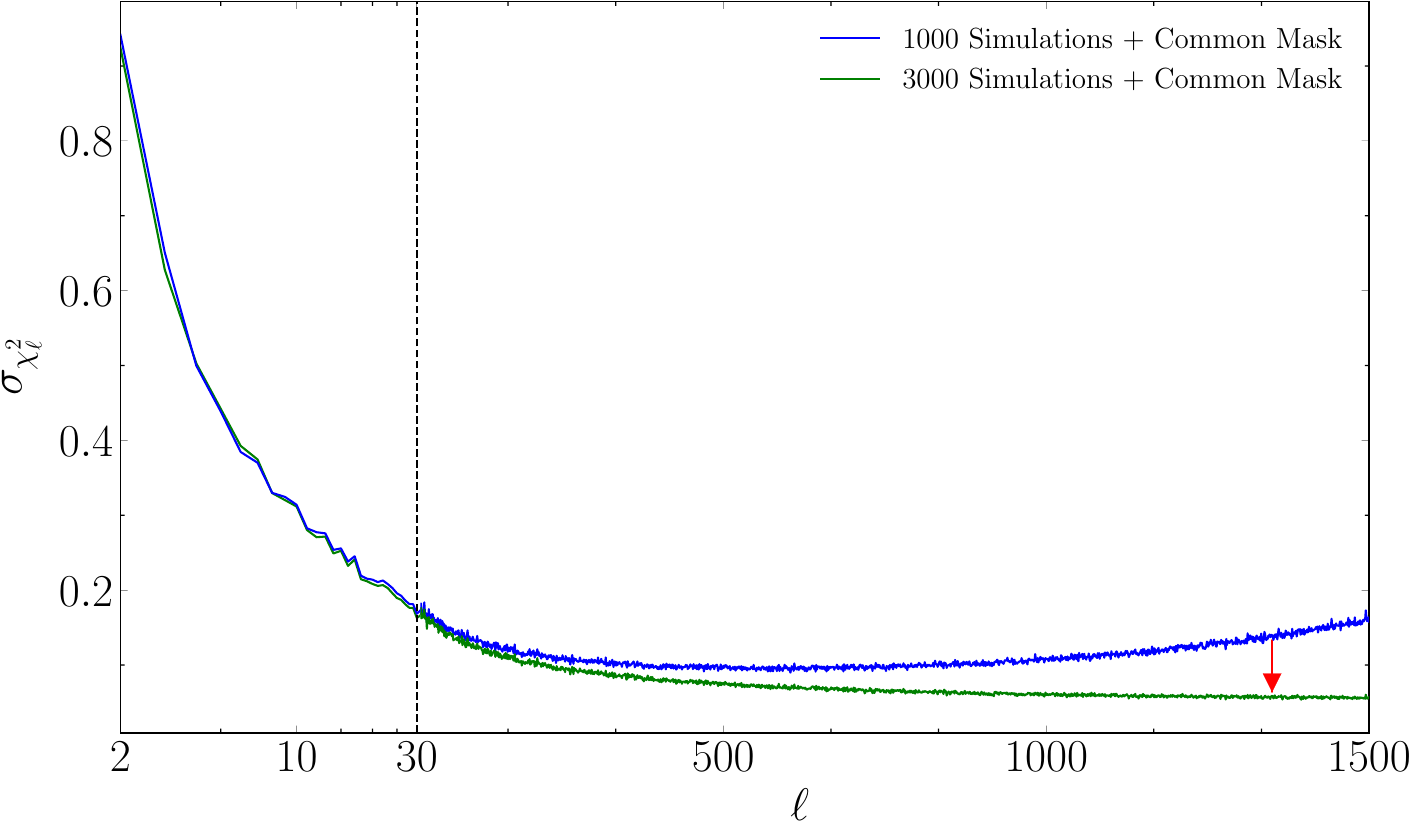}
    \caption{The standard-deviation of the cases of Fig.~\ref{fig:number-of-sims-mask}. Note that the values start to increase for $\ell > 500$ when we have only 1000 simulations, which indicate that numerical noise in the inversion of $C_{ij}$ starts to dominate. For 3000 simulations numerical noise is much suppressed.
    	\label{fig:variance-of-sims}
    }
\end{figure}

	The mean covariance matrix $C_{ij}$ appearing in Eq.~\eqref{chi-square} is estimated numerically. In practice, we would expect $C_{ij}$ to have converged to its theoretical value if $\chi^2_\ell$ approaches unity when applied to an independent masked sky. Figure~\ref{fig:number-of-sims-mask} shows $\chi^2_\ell$ as a function of $\ell$ for $C_{ij}$ evaluated with 1000 and 3000 simulations. Clearly, a higher number of independent maps is required to achieve $\chi^2_\ell\rightarrow1$ for all $\ell$s. As we have discussed in the main text, this is not a problem since the theoretical $\chi^2_\ell$ can be calibrated with the use of independent control simulations. However, the number of maps used to estimate $C_{ij}$ requires a more careful investigation since it will also have an impact on the variance of $\chi^2_\ell$, and consequently on the evaluation of the inverse correlation $(C^{-1})_{\ell_1\ell_2}$. This is illustrated explicitly in Figure~\ref{fig:variance-of-sims}, but can also be noted in the thickness of the variance bands on the high-$\ell$ tail of Figure~\ref{fig:number-of-sims-mask}. Note that the blue curve in Figure~\ref{fig:variance-of-sims} has a minimum at $\ell\gtrsim500$, which does not correspond to the expectation that cosmic variance should decrease with increasing $\ell$. With 3000 simulations this issue does not appear inside the range of scales we are probing, which indicates that the numerical noise is negligible in this case.
	
\section{Correlation among different multipoles}\label{app:corr-matrix}
\begin{figure}[t!]
    \centering
    \includegraphics[width = \columnwidth] {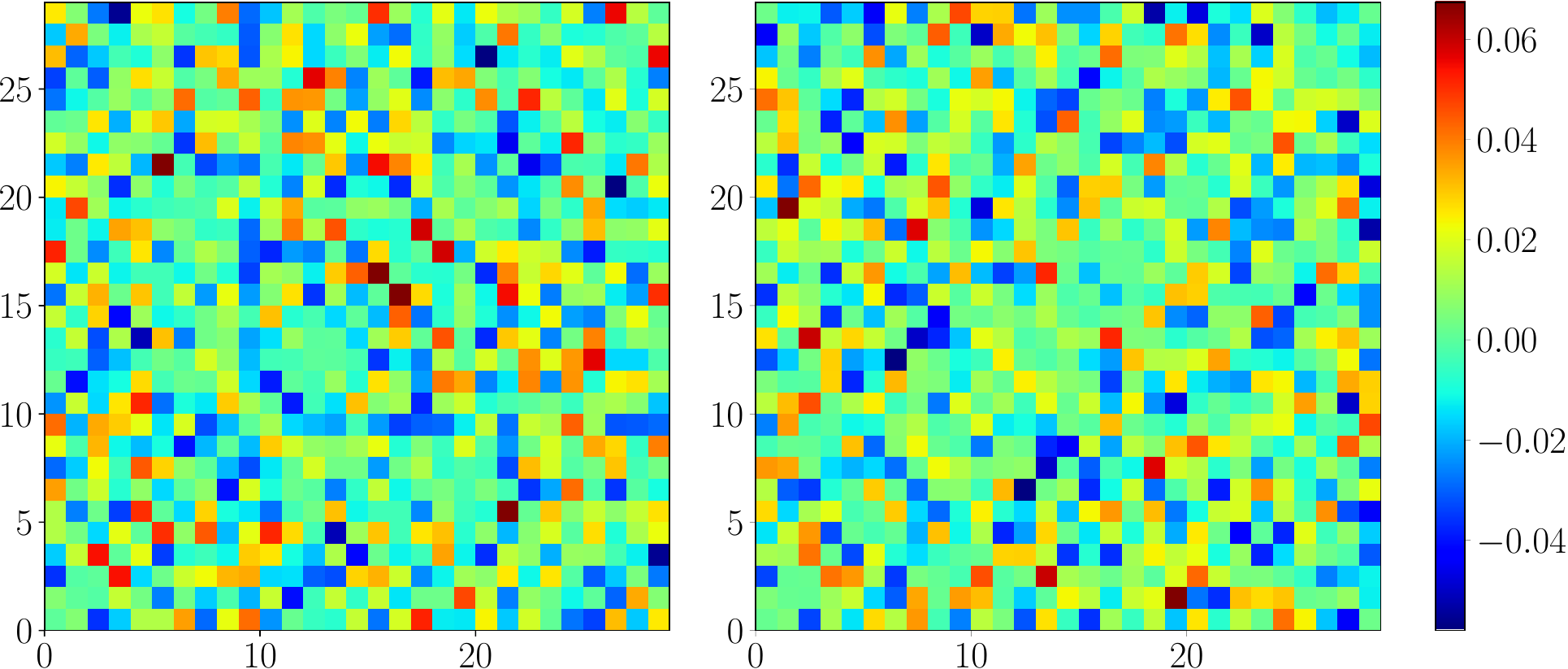}
    \caption{Matrix plot of the residual correlation matrix -- Eq.~\eqref{residual-corr} -- between $\chi^{2,\rm sim}_{\ell_1}$ and $\chi^{2,\rm sim}_{\ell_2}$. \emph{[Left]}: full sky maps; \emph{[Right]:} masked maps.\label{fig:corr-matrix}}
\end{figure}

The results of Tables~\ref{table-chi2-masked}--\ref{table-chi2-fullsky-2018} are based on the approximation $M_{\ell_1\ell_2}\propto\delta_{\ell_1\ell_2}$, where $M_{\ell_1\ell_2}$ is the covariance matrix whose inverse appears in Eq.~\eqref{eq:chisq-chisq} due to the presence of a mask (and possibly due to primordial non-Gaussianity). While this should be an excellent approximation at large multipoles, it was not guaranteed to hold at small $\ell$s. Figure~\ref{fig:corr-matrix} shows a matrix plot of the ``residual'' correlation matrix, defined as the correlation matrix minus the identity,
\begin{equation}\label{residual-corr}
    \rho_{\ell_1\ell_2} = \frac{M_{\ell_1\ell_2}}{\sigma_{\chi^2_{\ell_1}}\sigma_{\chi^2_{\ell_2}}} - \delta_{\ell_1\ell_2}\,,
\end{equation}
for the multipoles in the range $\ell\in[2,30]$. As we can see, different multipoles are weakly correlated even in these large scales, where the mask effect is strong. As a final test we computed the $\chi^2_{\rm MV}$ both with and without the non-diagonal correlations, and the results had negligible differences. We also computed the correlation matrix for the full range os scales here considered, and indeed the correlations seem to be negligible at all scales. This shows that the different multipoles can be treated as independent even in the presence of Planck's Common Mask.

\section{Choosing $N_{side}$}\label{app:nside}
	
\begin{figure}[t!]
    \centering
    \includegraphics[width =.95\columnwidth]
    {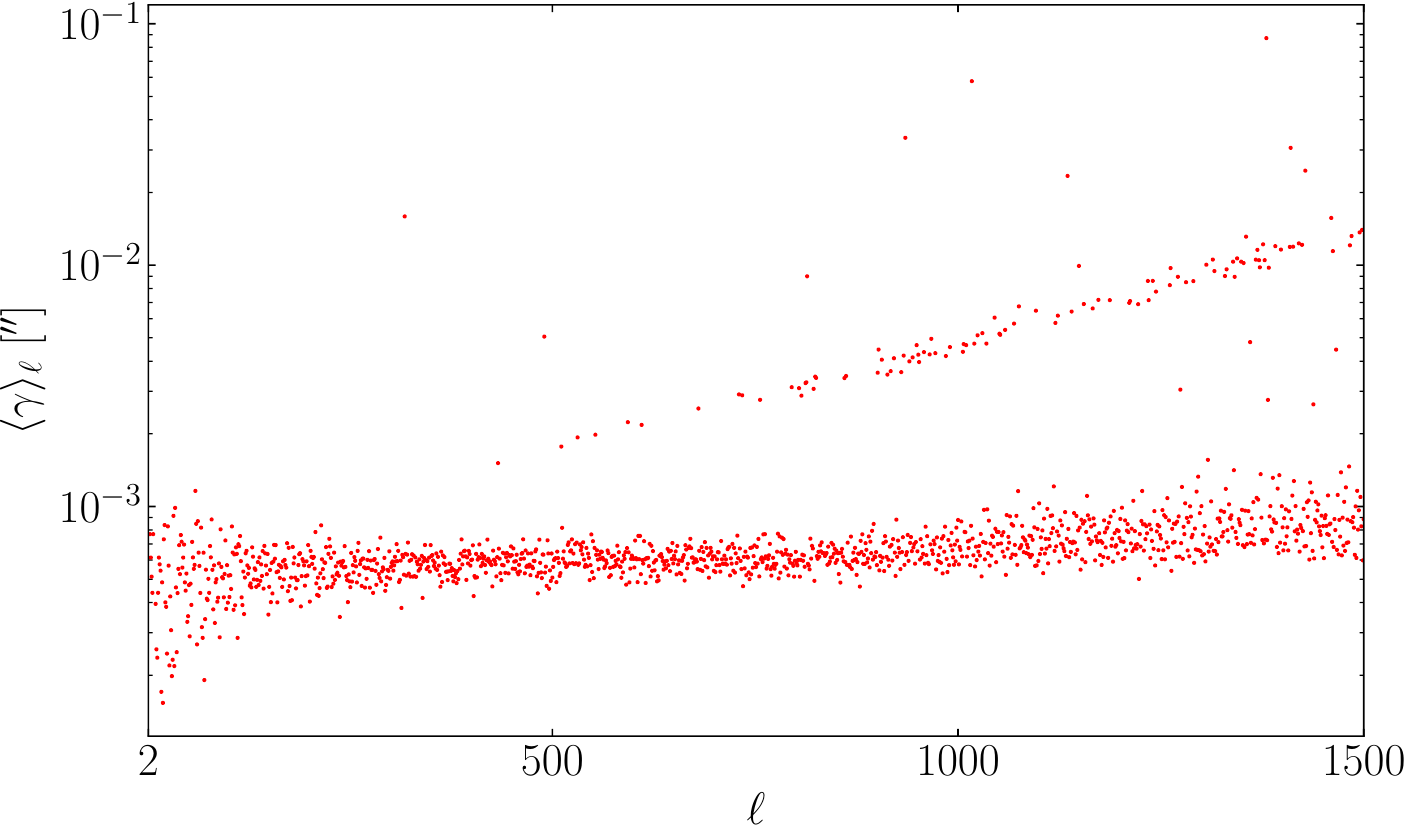}
    \caption{Mean angular displacement among MVs by changing resolution from $N_{side}=1024$ to $N_{side}=2048$ as a function of $\ell$. Both set of MVs were generated from CMB maps with the same random seed. The displacement of the MVs induced by a higher $N_{side}$ is of order of $10^{-3}\,{\rm arcsec}$, and thus negligible.}\label{fig:nside-choice}
\end{figure}

All MVs in this work were extracted from CMB maps constructed at resolution $N_{side}=1024$. Such resolution should enough to reconstruct maps at a maximum multipole ${\ell_{\rm{max}} = 3N_{side}-1\gtrsim 3000}$~\citep{Gorski:2004by} --- twice the maximum range used in this paper. Nonetheless, because numerical approximations in the reconstruction of the $a_{\ell m}$s is a potential source of systematic noise, which will in turn affect the position of the MVs, it is important to estimate the impact of choosing a higher $N_{side}$ in our simulations. Given a set of vectors at a fixed resolution, $\{\vecl{\ell}^{N_{side}}\}$, we can estimate this effect by evaluating the angle
\begin{equation}
	\gamma_{j,\ell} = \arccos(\vecl{j,\ell}^{2048}\cdot\vecl{j,\ell}^{1024})\,,
\end{equation}
with $\{\vecl{\ell}^{1024}\}$ and $\{\vecl{\ell}^{2048}\}$ sharing the same random seed --- and averaging it over all values of $j$. The mean angle ${\langle\gamma\rangle_\ell}$ is shown in Figure~\ref{fig:nside-choice}  as a function of $\ell$. The impact induced by a choice of a higher $N_{side}$ is below $10^{-3}\,{\rm arcsec} \sim 10^{-9}\,{\rm rad} $ for the vast majority of scales we probed. This corresponds to a multipole $\ell = \pi/\langle\gamma\rangle_\ell\sim 10^8$, which is five orders of magnitude above the maximum scale we are considering in this work. Even the highest observed difference of $0.1\,{\rm arcsec}$ corresponds to a multipole $\ell\sim10^6$, which is safely above our maximum range. Thus, the resolution $N_{side}=1024$ is completely safe for our purposes. Moreover, notice that the number of scales with a mean displacement $\gtrsim10^{-3}\,{\rm arcsec}$ increase linearly at $\ell\gtrsim 400$ --- roughly the same scale and shape above which ill-conditioning of the polynomials is observed, as confirmed by Figure~\ref{fig:timings}. Thus, such behavior in $\langle\gamma\rangle_\ell$ seen in Figure~\ref{fig:nside-choice} is most-likely due to numerical ill-conditioning of the polynomials, and not our choice of $N_{side}$.
	
\section{Inpainted maps}\label{app:inpainted}

\begin{figure*}[t!]
    \centering
    \includegraphics[width=0.7\columnwidth]{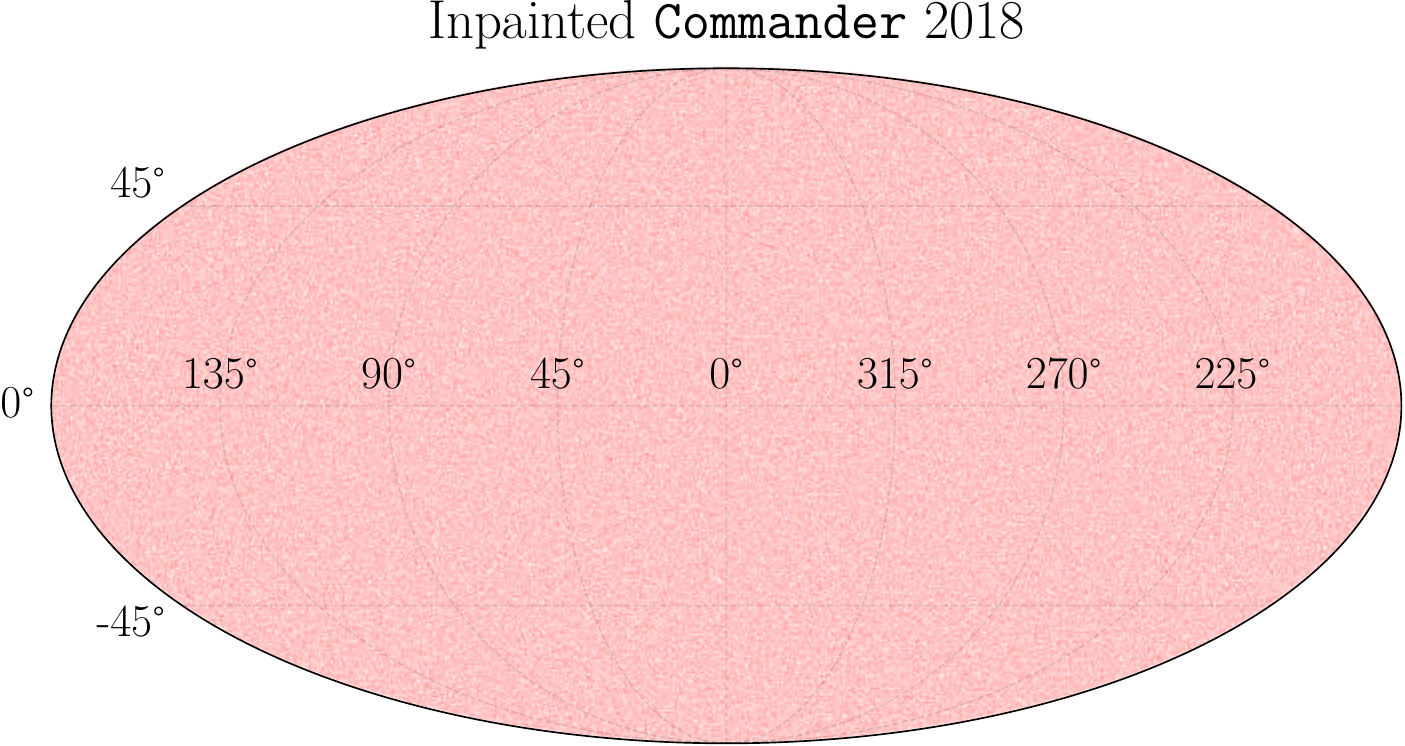}
    \qquad \includegraphics[width=0.7\columnwidth]{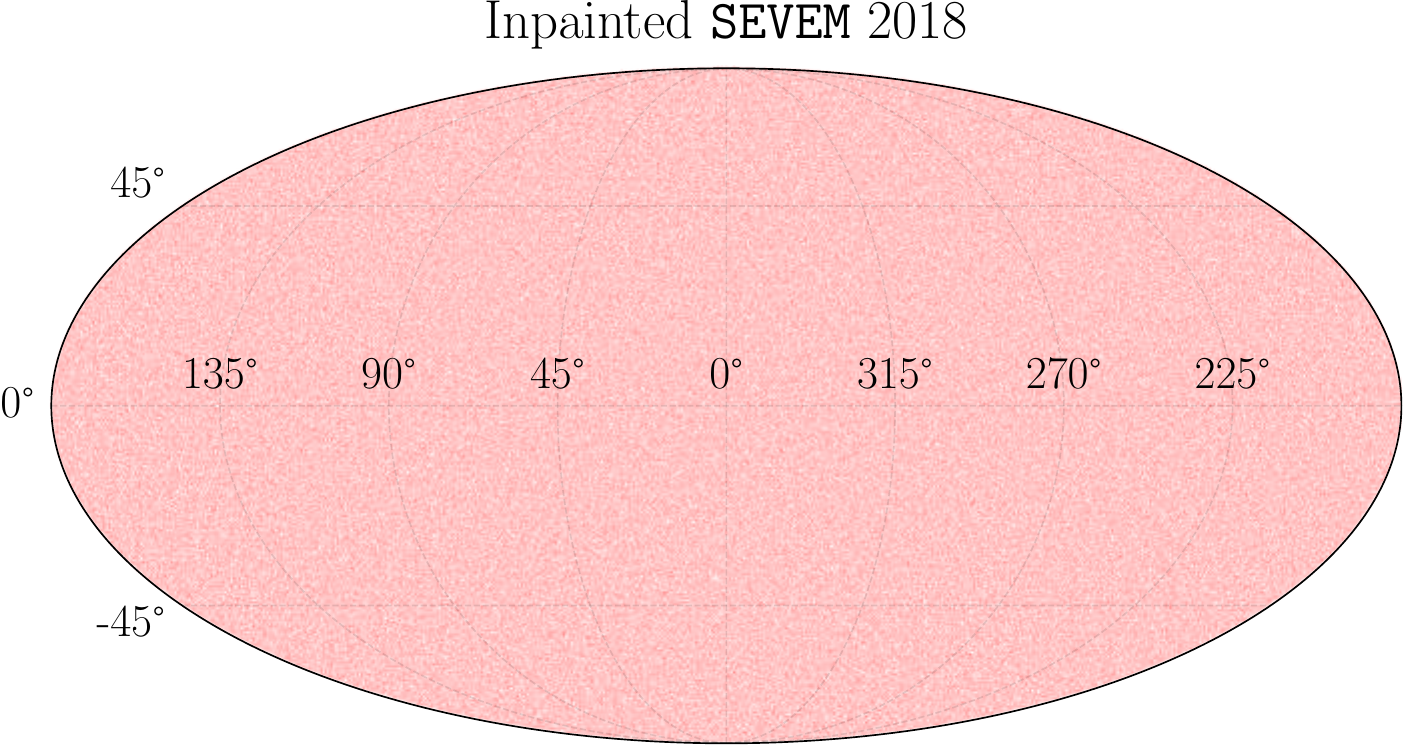}
    \caption{Similar to Figure~\ref{mollweide-mvs-fvs-fullsky} for the MVs of the 2018 {\tt Commander} and {\tt SEVEM} maps inpainted inside the small 2.1$\%$ mask. MVs for the inpainted {\tt NILC} and {\tt SMICA} maps are visually identical to the ones above, and thus not shown.\label{fig:inpainted-mvs}}	
\end{figure*}

Our analysis has shown that the 2018 {\tt Commander} and {\tt SEVEM} full-sky maps are clearly anisotropic in the range $\ell\in[2,1500]$. Since these maps contain anisotropic residuals which are visible by eye, mostly very near the galactic plane. We wanted to test if these visible residuals alone could account for the large $\sigma$-values reported in Table~\ref{table-chi2-fullsky-2018}. For this task we apply a simple inpainting method, which consists of masking these maps with the common inpainting mask made available by the Planck team (with $f_{sky}=0.979$), and then filling the masked regions with a realization of a Gaussian and statistically isotropic random sky. We used the same realization to inpaint all 4 maps. As we can see in Figure~\ref{fig:inpainted-mvs}, MVs for these inpainted maps are visually indistinguishable from full-sky Gaussian and isotropy random maps.

\begin{figure}[t!]
    \centering
    \includegraphics[width = .95\columnwidth] {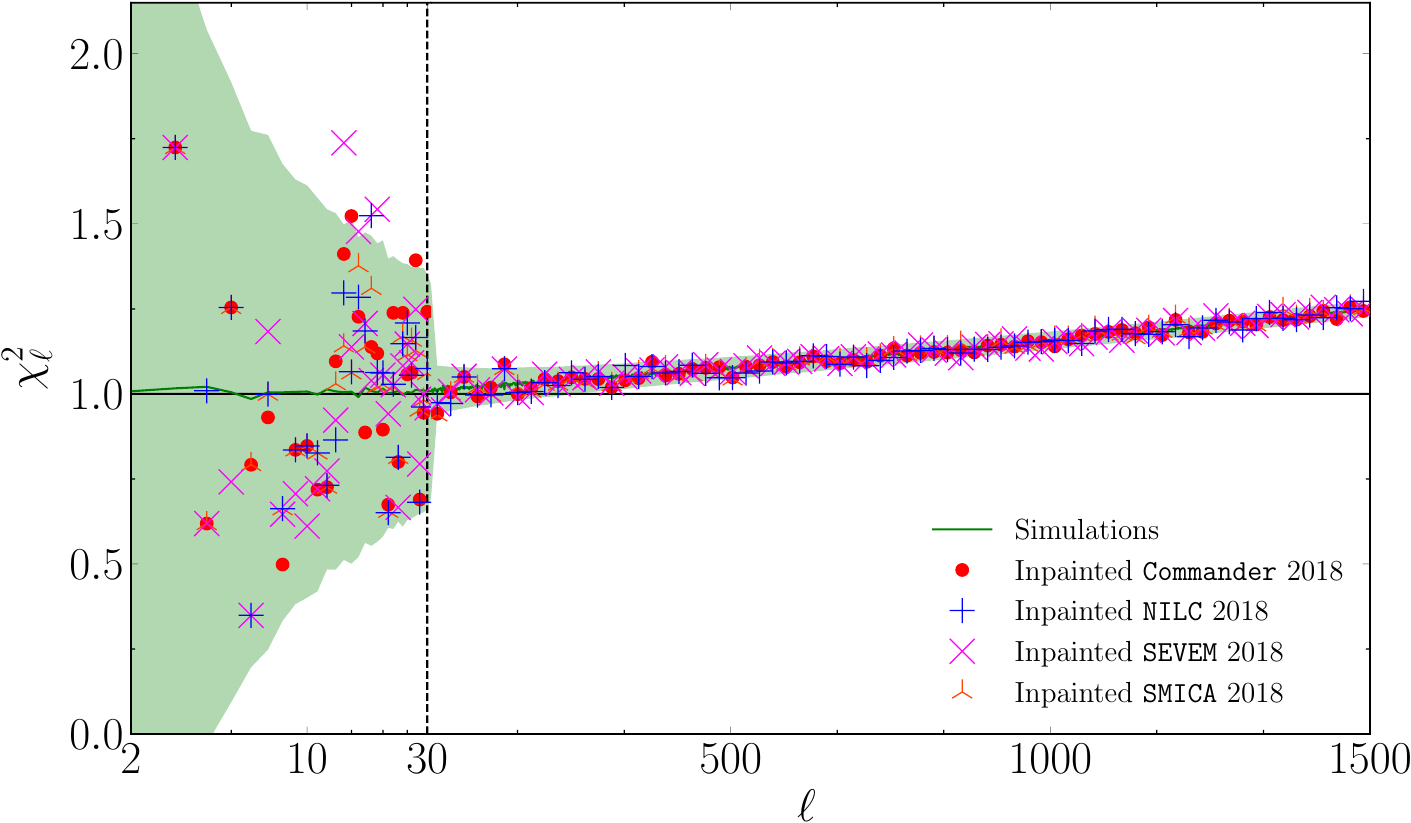}
    \caption{Same as Figure~\ref{masked-results} for the four inpainted Planck 2018 maps -- see the text for details. This small inpainting is enough to remove all anisotropies in our statistic. }
    \label{fig:inpainted-results}
    \medskip
    \footnotesize
    \setlength\tabcolsep{4pt}
    \begin{tabular}{cccccc}
        \toprule
        \multicolumn{2}{c}{\textbf{Full sky (PR3)}} & \!\!\texttt{Commander}\!\! & \texttt{NILC} & \texttt{SEVEM} & \texttt{SMICA}\\
        \midrule
        \midrule
        \multirow{2}{*}{Large scales} & $\chi^2_{\rm MV}/\text{dof}$ & 1.28 & 1.07 & 1.51 & 0.72\\
        \cmidrule{2-6} & $\sigma\text{-value}$ & \textbf{1.5} & \textbf{0.79} & \textbf{1.8} & \textbf{0.14}\\
        \midrule
        \multirow{2}{*}{WMAP scales} & $\chi^2_{\rm MV}/\text{dof}$ & 1.07 & 1.07 & 0.93 & 1.01 \\
        \cmidrule{2-6}
        & $\sigma\text{-value}$ & \textbf{1.6} & \textbf{1.7} & \textbf{0.17} & \textbf{0.79}\\
        \midrule
        \multirow{2}{*}{All scales} & $\chi^2_{\rm MV}/\text{dof}$ & 1.02 & 1.06 & 0.94 & 1.00 \\
        \cmidrule{2-6}
        & $\sigma\text{-value}$ & \textbf{1.0} & \textbf{1.9} & \textbf{0.08} & \textbf{0.70}\\
        \bottomrule
    \end{tabular}
    \captionof{table}{Same as Table~\ref{table-chi2-masked} for the data points of Figure~\ref{fig:inpainted-results}. \label{table-chi2-inpainted}}
\end{figure}

Figure~\ref{fig:inpainted-results} and Table~\ref{table-chi2-inpainted} show the result of the statistics~\eqref{chi-square-eta-plus-phi} applied to these maps, and Table summarizes the global fit of these maps when compare to full-sky Gaussian and isotropic simulations. This analysis shows that all anisotropies found in full-sky {\tt Commander} and {\tt SEVEM} maps are coming from this small 2.1\% fraction of the sky. Nevertheless, the fact is that the remaining 97.9\% of the sky is compatible with Gaussianity and isotropy, and this is an interesting and non-trivial result which confirms the predictions of the standard model to a great extent.

\vspace{6cm} 
\textcolor[rgb]{1.00,1.00,1.00}{.}

\textcolor[rgb]{1.00,1.00,1.00}{.}

\end{document}